\documentclass[12pt]{article}																																			  
\usepackage{bm, amsmath}
\begin{document}													     
\title{Theory of the liquid-glass transition in water}
\author{Toyoyuki Kitamura \\
Ecole Normale Sup\'{e}rieure de Lyon \\ 
 46, All\'{e}e d'Italie,
69364 Lyon Cedex 07,  France \\
Nagasaki Institute of Applied Science, Nagasaki 851-0193, Japan}


\maketitle
\vspace{1.5 cm}
\begin{abstract}
A quantum field theory of the liquid-glass transition in water based on
the two band model in the harmonic potential approximation is presented
by taking into account of the hydrogen bonding effect and the 
polarization effect. The sound and diffusion
associated with intra-band density fluctuations, and the phonons and
viscocity associated with inter-band density fluctuations are calculated.
The Kauzmann paradox on the Kauzmann's entropy
crisis and the Vogel-Tamman-Fulcher (VTF) law on the relaxation times
and the transport coefficients are elucidated from the sound instability
at a reciprocal particle distance corresponding  a hydrogen bond
length and at the sound instability temperature very
close to the Kauzmann temperature. The gap of specific heat at the glass
transition temperature and the boson peaks are also
presented. 
\end{abstract}

\vspace{5 cm}
\noindent
{\bf Pacs codes}: 63.10.+a, 64.70.Pf. \\
{\bf Key words}: water, cluster, glass transition, Kauzmann paradox,
Vogel-Tamman-Fulcher law, boson peak.

\newpage
\section{Introduction}
\hspace{5ex}
Water plays an essential role in a medium in which  chemical
and biological molecules operate and it influences their
functions. Its properties are far from being fully understood.
Experimental studies of the glassy behaviours of bulk water
are extremely difficult, while hydration water particularly
on the surface of protains can provides an interesting
alternative to study  the slow dynamics of water, because
the binding of water molecules to the surface can prevent
crystallization. Experimental studies in hydration water
certainly show the universal features in the liquid-glass
transition \cite{Careri,MP,Halle,Russo}. We intend to
develop theory of the liquid-glass in hydration water.
Here at the first
stage we develop the glass transition in bulk water.

We have established quantum field theory of the
liquid-glass transition in a one-component liquid in
the framework of the two band model to elucidate the
Kauzmann paradox, the VTF law,  the gap of specific
heat at the glass transition temperature and boson peaks
\cite{K1,K2}, which appear universally in the liquid-glass
transition. We have also proposed a quantum field theory
of  phonons in a two-component liquid \cite{K1,K3,K4},
and  also proposed a quantum field theory of
phonons and melting in compounds
taking into account of the electric
polarization in the rigid ion approximation,
the Ewald method \cite{K5,Mar,Ewa}.
Here we extend the quantum field theory of
the liquid-glass transition in a one-component liquid
to that in water considering
the quantum field theory of phonons in a multi-component
liquid and the electric polarization in the rigid ion
approximation \cite{K5,Mar,Ewa}. 
The essential points of our
theory are as follows \cite{K1, K2}:

(1) The two band model. A particle is temporarily in a  randomly
distributed harmonic potential made up by the surrounding
particles, making up and down transitions between the ground
state and the first excited states, and then hops to a surrounding
vacancy. This picture constitutes the model Hamiltonian
which consists of the random eigenfrequencies and
random hopping magnitudes. The configurational averaged
model Hamiltonian yields an unperturbed Hamiltonian.
Due to hopping, instead of individual states of the local
potentials, the system have energy bands having a width
determined by the hopping magnitude.

(2) Intra-band and inter-band density fluctuations. Density
fluctuations consist of intra-band density fluctuations
related to the hopping to a neighbouring vacancy
and inter-band density
fluctuations associated to the up and down transitions
between two bands.  Intra-band and inter-band density
fluctuations are associated with  intra-band and
inter-band elementary excitations, respectively.
The collective excitations of intra-band and inter-band
elementary excitations are sound and phonons, respectively.
Sound is a density fluctuation wave. In the classical limit
the particle energy dispersion associated to intra-band
elementary excitations tends to a free particle one
so that the sound reduces to a density fluctuation
wave \cite{K6}.
Phonons are elastic waves. In the classical limit coupling
strengths between particles deviated from their assigned
positions constitute spring constants so that the phonons
reduce to  elastic waves \cite{K6}.

(3) Dissipation and relaxation times. Random scattering
processes come from the random potentials which
is described by subtracting the unperturbed Hamiltonian
from the model Hamiltonian. Dissipative process comes
from the simultaneous scattering processes of
two particles by the same random potential. The sum of
the configurationally averaged simultaneous scattering
processes over all sites yields the correlation functions of
random harmonic frequencies and random hopping magnitudes,
which leads to the relaxation times; the former and the latter
correspond to the $\beta$ and $\alpha$-relaxations,
respectively. When the relaxation times of sound and phonons
are shorter than their respective periods, diffusivity
and viscosity appear, respectively.

(4) The Kauzmann entropy and the VTF law. Intra-band density
fluctuation entropy consists of sound entropy, fluctuation
entropy due to intra-band elementary excitations and
dissipative entropy due to diffusion. In an equilibrium state
the dissipative entropy compensates the fluctuation entropy
with a negative value and the system reaches a local
equilibrium. But since sound depends on
temperature, a mixing between the sound and fluctuation
entropies occurs. The mixing entropy yields the Kauzmann
entropy: a Curie law with a minus sign, where the critical
temperature corresponds to the sound
instability temperature which indicates fragility.
The hopping amplitude relates to the multi-dimensional configuration
space of  the Kauzmann entropy; the hopping amplitude
is proportional to the exponent of the Kauzmann entropy
per particle. This hopping amplitude explains
the VTF law: the exponent of the Curie law with a minus
sign per particle. The hopping amplitude is the origin of
the $\alpha$-relaxation.

(5) The glass transition temperature and the magnitude of
the randomness of harmonic frequencies. Inter-band density 
fluctuation entropy also consists of phonon entropy, fluctuation
entropy due to inter-band elementary excitations and
dissipative entropy due to viscosity. In an equilibrium
state the dissipative entropy compensates the fluctuation
entropy with a negative value and the system reaches
a local equilibrium. Since phonons do not depend
on temperature, mixing does not occur. 
The magnitude of randomness of harmonic frequencies relates
to the multi-dimensional configurational space of the
fluctuation entropy; the magnitude of randomness of
harmonic frequencies is
proportional to the exponent of the fluctuation entropy
per particle. The fluctuation entropy prevents the
Kauzmann entropy crisis at the glass transition temperature.
The magnitude of the randomness of harmonic
frequencies is the origin of the $\beta$-relaxation.

In applying the two band model to the liquid-glass transition
in water \cite{Eis,Han,Hey}, we should clarify the structure of water.
The characteristic features of the structure of water are as follows
\cite{Sta}: 
(i) Hydrogen
bonds play an essential role in clustering water molecules comparing
with the common molecules.  A water molecule consists of  an oxygen
and two covalent bonded hydrogens so that the water molecule
has  the two protons and the lone pairs. A proton of a water molecule
is hydrogen bonded to a lone pair of  the nearest neighbour water
molecule.
(ii) Water has the proper geometric structure. Water is made up of
a strongly hydrogen bonded network and locally very similar to ice.
But these hydrogen bonds have very short lifetimes. 
(iii) A water molecule is electrically polarized, which originates from
the hydrogen covalent bonds.

In order to include the above characteristic features, first we determine
the interaction potentials $V_{\alpha\beta}$ with
the corresponding pair distribution functions $g_{\alpha\beta}$,
which predict the structure of water, where $\alpha$, $\beta$
mean an oxygen atom O or a hydrogen atom H. 
(i)  The interaction potential between O-H,  $V_{\rm OH}$ consists
of  hydrogen covalent bonding,  $V_{\rm OH}^{\rm hcb}$  and
hydrogen bonding, $V_{\rm OH}^{\rm hb}$. A strong hydrogen
bonded network is constructed by  the couplings of
$V_{\rm OH}^{\rm hcb}V_{\rm HO}^{\rm hb}$ and
$V_{\rm OH}^{\rm hb}V_{\rm HO}^{\rm hcb}$, which comes from the fact
that the configurational  average of  the term
$V_{\rm OH}V_{\rm HO}$
over the position of the hydrogen atoms
yields the terms $V_{\rm OH}^{\rm hcb}V_{\rm HO}^{\rm hb}$ and
$V_{\rm OH}^{\rm hb}V_{\rm HO}^{\rm hcb}$, not the terms
$V_{\rm OH}^{\rm hcb}V_{\rm HO}^{\rm hcb}$
and $V_{\rm OH}^{\rm hb}V_{\rm HO}^{\rm hb}$, because the
former terms are favorable.
The hydrogen bonding plays an essential
role in clustering water.  The first peak of $g_{\rm OH}(R)$
 corresponds to the hydrogen bond length \cite{Hey}.  Since hydrogen
bonding plays an essential role in water, 
the sound instability occurs at a reciprocal
particle distance corresponding to a hydrogen bond length.
(ii)  In the interaction potential $V_{\rm HH}$ we take into
account of the interaction between hydrogens in different water
molecules. Hopping between  protons  and lone pairs reflects
$V_{\rm HH}$
and the hopping yields very short lifetimes of hydrogen bondings.
(iii) There exists the electric polarization between O-H. The electric
polarization relates to the $p$-state of particles.
An excited state of a particle corresponds to
a $p$-state, while a ground state to an $s$-state. Thus
we need not take into account of the polarization in the interaction
potentials between intra-band density fluctuations because the
intra-band density fluctuations annihilate and create the same
states. However we should take into account of the polarization
between inter-band density fluctuations because
the inter-band density fluctuations annihilate an  $s$-state
and create a $p$-state, and vice versa.  We take into account
of the electric polarization in the rigid ion approximation,
the Ewald method \cite{K5,Mar,Ewa}.

We organize as follows:
in section 2,  we briefly survey the formulation. 
We obtain the model Hamiltonian with random harmonic
frequencies and random hopping magnitudes. Taking
the configurational average of the model Hamiltonian,
We obtain the
unpurterbed Hamiltonian. We construct the interaction
Hamiltonian for intra-band and inter-band density fluctuations,
which leads to the sound and the phonon modes.
In section 3, we introduce the interaction Hamiltonian for the
random scattering processes, which leads to the dissipative
process.
In section 4, the correlation fuctions of intra-band density
fluctuations with the vertex corrected
bubble diagrams of intra-band elementary excitations
yield the sound and diffusivity. The sound instability occurs at 
a temperature $T_0$ and at a reciprocal particle distance corresponding
to a hydrogen bond length. In section 5, taking into account of 
the electric polarization due to hyrogen
covalent bond, we calculate the correlation functions of inter-band
density fluctuations with the vertex corrected bubble diagrams,
which yield phonons and viscosity modified by the polarization.
We also show the boson peaks.
In section 6,  we obtain the Kauzmann entropy from the mixing of
the sound and the intra-band fluctuation entropies in the sound
instability  regime. We show that
a unique hopping amplitude corresponding to the Kauzmann
entropy governs the velocities of  individual particles, the sound
and the hopping amplitudes of individual particles,
while the magnitude of the randomness
of the hamonic frequencies of particles depends
on the individual particles. The glass transition occurs at the crossover
temperature of the Kauzmann entropy and the intrer-band
fluctuation entropy. We derive the specific heat and show 
the VTF law for the relaxation times for sound and phonons, and
the transport coefficients. Section 7 is devoted by some concluding
remarks.

\section{Formulation}
\hspace{5ex}
Here we briefly survey the formulation.
We start with the following Hamiltonian:
\begin{equation}
H=\sum_\alpha \int\,d^3x
\psi_\alpha^\dagger(x)\hbar\epsilon_\alpha^0
(-{\rm i}\bm{\nabla})\psi_\alpha(x)+
\frac12\sum_{\alpha\beta}\int\,d^3xd^3y
n_\alpha(x)V_{\alpha\beta}
(\bm{ x}-\bm{y})	n_\beta(y),
\tag{2.1}
\end{equation}
where $\hbar\epsilon_\alpha^0(-{\rm i}\bm{\nabla})$
is the energy operator
of a free $\alpha$-particle,  $n_\alpha(x)=
\psi_\alpha^\dagger(x)\psi_\alpha(x)$.
We write the operator $\psi_\alpha(x)$ by using
the localized operators,
which are rewritten by the extended operaters.
\begin{equation}
\psi_\alpha(x)=\sum_{m \mu} 
\tilde w_{\alpha\mu}(\bm{ x}-\bm{R}_{\alpha m})
b_{\alpha m \mu }.
\qquad
b_{\alpha m \mu}={1 \over \sqrt{N_\alpha}}\sum_{\bm p}
e^{{\rm i}\bm{ p\cdot R}_{\alpha m}}a_{\alpha\mu {\bm p}}.
\tag{2.2}
\end{equation}
The Heisenberg equation for $b_{\alpha m\mu}$ is given by
\begin{equation}
-\hbar {\partial \over \partial \tau}b_{\alpha m\mu}=
\sum_{n \nu}\int\,
d^3x\tilde w_{\alpha m\mu}(\bm{ x}-\bm{R}_{\alpha m})
\left[\hbar
\epsilon^0_{\alpha}
(-{\rm i}\nabla)+ \Phi_{{\bm R}_{\alpha n}}({\bm x})\right]
\tilde w_{\alpha n\nu}(\bm{ x}-\bm{R}_{\alpha n})
b_{\alpha n\nu},
\tag{2.3}
\end{equation}
\begin{equation}
\Phi_{{\bm R}_{\alpha n}}({\bm x})=\sum_\beta 
\int_{\{{\bm R}_{\alpha n}\}}
\, d^3y V_{\alpha \beta}(\bm{ x}-\bm{y})	n_\beta(y),
\tag{2.4}
\end{equation}
where $\{{\bm R}_{\alpha n}\}$ means summing the surrounding
sites of the particle at  
${\bm R}_{\alpha n}$ in $n_\beta(y)$.
The term $\Phi_{{\bm R}_{\alpha n}}$ means that a 
potential is determined by the surrounding 
particles of the site ${\bm R}_{\alpha n}$. Thus we obtain the
Schr\"{o}dinger	equation:
\begin{equation}
\left\{\hbar \epsilon^0_\alpha(-{\rm i}\nabla)+
\Phi_{{\bm R}_{\alpha n}}({\bm x})\right\}
\tilde w_{\alpha n\nu}
(\bm{ x}-\bm{R}_{\alpha n})=
\hbar \tilde \omega_{\alpha n\nu
}\tilde w_{\alpha n\nu}(\bm{ x}-\bm{R}_{\alpha n}),
\tag{2.5}
\end{equation}
where $\tilde \omega_{\alpha n\mu}$ and
$\tilde w_{\alpha n\mu}$
are the eigenfrequency
and	the eigenfunction of the $\mu$th state at the site
${\bm R}_{\alpha n}$ in a simmilar manner to a one-component liquid
in the harmonic approximation.
We consider the following relations:
\begin{equation}
w_{n\alpha1}(x_i)=-2\zeta_{n \alpha} \nabla_i w_{n\alpha0}(x_i);
\quad
w_{n\alpha0}(x_i)=2\zeta_{n \alpha} \nabla_i w_{n\alpha0}(x_i);
\qquad \zeta_{n \alpha}=
\sqrt{\frac{\hbar}{2M_\alpha\omega_{n \alpha}}},
\tag{2.6}
\end{equation}
where $\zeta_{n \alpha}$ is the mean width of the zero point motion,
$M_\alpha$ is the mass of the $\alpha$-particle and
$\omega_{n \alpha}$ is the harmonic frequency; 
$\hbar\tilde \omega_{n\alpha0}=\frac32\hbar\omega_{n\alpha}$ and 
$\hbar\tilde \omega_{n\alpha i}=\frac52\hbar\omega_{n\alpha}$.
At this stage, we can take  the 
following model	Hamiltonian in terms of localized operators:
\begin{equation}
H=\sum_{\alpha m\mu}\hbar \tilde \omega_{\alpha m\mu}
b_{\alpha m\mu}^\dagger	b_{\alpha m\mu}
+\sum_{\alpha m\mu n\nu}\hbar J_{\alpha m\mu \alpha n\nu}
b_{\alpha m\mu}^\dagger b_{\alpha n\nu},
\tag{2.7}
\end{equation}
where the hopping matrix is given by
\begin{equation}
\hbar J_{\alpha m\mu \alpha n\nu}=\int\,d^3x
\tilde w_{\alpha m\mu}(\bm{ x}-\bm{R}_{\alpha m})
\Phi_{{\bm R}_{\alpha n}}({\bm x})
\tilde w_{\alpha n\nu}(\bm{ x}-\bm{R}_{\alpha n}).
\tag{2.8}
\end{equation}
Note that the interaction Hamiltonian is nonlinear through the term
$n_\beta(y)$ 
in $\Phi_{{\bm R}_{\alpha n}}$ and the potential
$\Phi_{{\bm R}_{\alpha n}}$ permits a particle
to hop only to a vacancy.

Now we start with the new unperturbed Hamiltonian which is the
configurational average of the model Hamiltonian (2.7):
\begin{equation}
H_0= <H>_c=\sum_{\alpha \mu m}\hbar \tilde \omega_{\alpha \mu} 
b^\dagger_{\alpha m\mu}	b_{\alpha m\mu}
+\sum_{\alpha \mu mn}\hbar J_{\alpha \mu}
({\bm R}_{\alpha m}-{\bm R}_{\alpha n})
b^\dagger_{\alpha m\mu}b_{\alpha n\mu},
\tag{2.9}
\end{equation}
where $J_{\alpha\mu} $ is the hopping magnitude.
$<\ >_c$ means the configurational average.
The configurationally
averaged values do not include the indices of the positions.
Here for simplicity we limit
ourselves to the hopping between like atoms and 
the hopping to those between the same levels. Using Eq.(2.2), we 
obtain
\begin{equation}
H_0=\sum_{\alpha\mu{\bm p}} \hbar\epsilon_{\alpha\mu{\bm p}}
a^\dagger_{\alpha\mu{\bm p}}
a_{\alpha\mu{\bm p}},
\qquad
\epsilon_{\alpha\mu{\bm p}}=\tilde \omega_{\alpha\mu}+J_{\alpha \mu}
({\bm p}),
\qquad
J_{\alpha\mu}({\bm p})\equiv \frac1{N_\alpha} \sum_m e^{i\bm{ p\cdot 
R}_{\alpha m}}J_{\alpha \mu}({\bm R}_{\alpha m}).
\tag{2.10}
\end{equation}
From Eq.(2.2), we obtain 
\begin{equation}
n_\alpha(x)=\frac1{N_\alpha}\sum_{m{\bm q}}
e^{-i\bm{ q\cdot R}_{\alpha m}}
[\sum_\mu \tilde w^2_{\alpha\mu}(
\bm{ x}-\bm{R}_{\alpha m})\rho^\dagger_{\alpha d\mu{\bm q}}-\sum_i 
\zeta_\alpha\nabla_i
\tilde w^2_{\alpha0}(\bm{ x}-\bm{R}_{\alpha m})
\rho^\dagger_{\alpha i{\bm q}}
+\cdots],
\tag{2.11}
\end{equation}
where
\begin{equation}
\rho_{\alpha d\mu{\bm q}}=\sum_{\bm p}
a^\dagger_{\alpha\mu{\bm p}}a_{\alpha\mu\bm{ p+q}},
\qquad
\rho_{\alpha i{\bm q}}=\sum_{\bm p}(a^\dagger_{\alpha i{\bm p}}
a_{\alpha0\bm{ p}+\bm{q}}+
a^\dagger_{\alpha0{\bm p}}a_{\alpha i\bm{ p}+\bm{q}}).
\tag{2.12}
\end{equation}
$\rho_{\alpha d\mu{\bf q}}$ and
$\rho_{\alpha i{\bf q}}$
are intra-band and inter-band density fluctuation operators, respectively. 
Substituting Eq.(2.11) into the interaction Hamiltonian in Eq.(2.1) we
obtain the interaction Hamiltonian:
\begin{equation}
H_I=\frac12	{\sum_{\alpha \beta\mu \nu {\bm q}}}'
\frac1{\sqrt{N_\alpha N_\beta}}
V^{d}_{\alpha \mu\beta\nu}({\bm q})\rho^\dagger_{\alpha d
\mu{\bm q}}\rho_{\beta d\nu{\bm q}}
+\frac12 \sum_{\alpha \beta ij{\bm q}}
\frac1{\sqrt{N_\alpha N_\beta}}
V^{od}_{\alpha i\beta j}({\bm q})\rho^\dagger_{\alpha i{\bm q}}
\rho_{\beta j{\bm q}},
\tag{2.13}
\end{equation}
\begin{equation}
V^d_{\alpha \mu \beta \nu}({\bm q})\equiv \sqrt{N_\alpha N_\beta}
V_{\alpha\mu\mu,\beta\nu\nu},
\qquad V^{od}_{\alpha i\beta j}({\bm q})
\equiv \sqrt{N_\alpha N_\beta}V_{\alpha 0i,\beta 0j},
\tag{2.14}
\end{equation}
\begin{multline}
V_{\alpha\mu\mu',\beta\nu\nu'}({\bm q})=
\frac1{N_\alpha N_\beta}
\sum_{m\ne n}e^{-i{\bm q}\cdot
({\bm R}_{\alpha m}-{\bm R}_{\beta n})} \\
\times\int\,d^3xd^3y
\tilde w_{\alpha\mu}(\bm{ x}-\bm{R}_{\alpha m})
\tilde w_{\alpha\mu'}(\bm{ x}-\bm{R}_{\alpha m})
V_{\alpha\beta}(\bm{ x}-\bm{y})
\tilde w_{\beta\nu}(\bm{ y}-\bm{R}_{\beta n})
\tilde w_{\beta\nu'}(\bm{ y}-\bm{R}_{\beta n}),
\tag{2.15}
\end{multline}
where the prime on $\Sigma$ means $H_I$ excludes loop diagrams, which
are taken into accout in the potential
$\Phi_{{\bm R}_{\alpha n}}$.

\section{Dissipative process}
\hspace*{5ex}
In randomly distributed particles, we have a model Hamiltonian,
Eq. (2.7). Taking the configurationally averaged
Hamiltonian of Eq.(2.7), we obtain the unperturbed Hamiltonian,
Eq.(2.9) and (2.10), which 
constitutes the two bands. The intra-band and inter-band elementary
excitations are excited around the two bands and they are associated
with intra-band and inter-band density fluctuations, Eq.(2.12), which
constitute the model Hamiltonian (2.13). We employ the model
Hamiltonian (2.13) to describe
the dynamical process. The collective excitations for the intra-band
and inter-band elementary excitaions are sound and phonons. The life
times of sound and phonons appears when the dispersion curves merge
into the continuum of the intra-band and inter-band elementary
excitations, respectively.

Here we investigate the dissipative process from the random scatterings
due to the random harmonic frequencies and the random hopping magnitudes.
The original model Hamiltonian, Eq.(2.7) reflects the randomly
distributed particles, while the unperturbed Hamiltonian, Eqs.(2.9) and
(2.10) reflect the free particle picture. The random scattering 
process is represented by the interaction Hamiltonian $H_I$:
$$
H_I=\sum_{\alpha m \mu} 
\hbar (\tilde \omega_{\alpha m \mu }-
\tilde \omega_{\alpha \mu})
b_{\alpha m \mu}^\dagger
b_{\alpha m \mu}+\sum_{\alpha mn \mu}\hbar 
\{J_{\alpha m\mu  \alpha n\mu }-
J_{\alpha \mu}({\bm R}_{\alpha m}-{\bm R}_{\alpha n})
\}b_{\alpha m \mu}^\dagger b_{\alpha n \mu}.
\eqno(3.1)
$$
Note that since the term $J_{\alpha m\mu \alpha n\mu}$
 involves operators $n(y)$,
$H_I$ is essentially nonlinear. But at low temperatures an atom stays 
longer at the same  site. Thus we make the approximation that the term
$J_{\alpha m\mu \alpha n\mu}$ is a  
random c-number. Using Eq.(2.2), we rewrite Eq.(3.1) as
\begin{multline}
H_I= \frac1{N_\alpha} \sum_{\alpha \mu \bm{ pp'}}\left[\sum_m\hbar
(\tilde \omega_{\alpha \mu m}-
\tilde \omega_{\alpha \mu})
e^{-{\rm i}(\bm{ p}-\bm{p'})\cdot 
{\bm R}_{\alpha m} } \right. \\
 +\left.\sum_{mn} 
\hbar\{J_{\alpha m\mu \alpha n\mu}-
J_{\alpha \mu}({\bm R}_{\alpha m}
-{\bm R}_{\alpha n})\}
e^{-{\rm i}\bm{ p\cdot R}_{\alpha m}+
{\rm i}\bm{ p'\cdot R}_{\alpha n} }\right]
a_{\alpha \mu{\bm p}}^\dagger
a_{\alpha \mu {\bm p'}}.
\tag{3.2}
\end{multline}

Dissipative process is constructed by elementary scattering processes
due to simultaneously scattering processes of two particles in
$\mu$ and $\nu$th bands by the same random potentials. Elementary
scattering processes come from the scattering processes due to
random harmonic frequencies and random hopping magnitudes, Eqs.(3.1)
and (3.2). The sum of configurationally averaged
elementary scattering process over all
sites due to random harmonic frequncies yields a correlation functions
of random harmonic frequencies, $U^{\alpha \mu\nu}_\omega$ and that due to
random hopping magnitudes yields a correlation function of random
hopping magnitudes, $U^{\alpha \mu\nu}_J$, which are given by  
$$
U^{\alpha \mu\nu}_\omega=\frac1{N^2_\alpha}\sum_m
(<\tilde \omega_{\alpha \mu m}
\tilde \omega_{\alpha \nu n}>_c-
\tilde \omega_{\alpha \mu} \tilde \omega_{\alpha \nu}),
\eqno(3.3)
$$
$$
U^{\alpha \mu\nu}_J(q)=\frac1{N_\alpha}
\int\,d^3Rg(R)e^{-{\rm i}\bm{ q\cdot R}}
\left. \{<J_{\alpha m\mu n\mu} J_{\alpha m\nu n\nu}>_c-
J_{\alpha \mu}({\bm R}) J_{\alpha \nu}({\bm R})\}
\right\vert_{{\bm R}=
{\bm R}_{\alpha m}-{\bm R}_{\alpha n} }. 
\eqno(3.4)
$$
Eqs.(3.3, 3.4) yield the relaxation times; the $\beta$ and
$\alpha$-relaxation, respectively. The elementary scattering
processes, $U^{\alpha \mu\nu}_{\omega, J}$ correspond to
that in the case of  two electrons scattered
simultaneously by the same random
impurity in electric conductivity. The statistical
average on a random system constrains the dynamical
processes, which leads to dissipation.

\section{Sound and diffusivity}
\hspace{5ex}
We first investigate the sound Green's function.
In the low temperature regime, we confine ourselves to the lower band.
$$
F_{\alpha\beta{\bm q}}(\tau_1-\tau_2)\equiv-
\frac1{\hbar\sqrt{
N_\alpha N_\beta}}<T_\tau\rho_{\alpha 0{\bm q}}(\tau_1)
\rho^\dagger_{\beta 0 {\bm q}}(\tau_2)>_c \equiv
\frac1{\beta\hbar}\sum_{{\rm i}\nu_n}
e^{-{\rm i}\nu_n(\tau_1 -\tau_2)}
F_{\alpha \beta}(q),
\eqno(4.1)
$$
where we should not confuse the temperature
$\beta=1/k_BT$ and
the suffices $\beta$; the temperature $\beta$ appears as an
independent parameter with the inverse energy dimension, while
the suffices $\beta$ stand for  O or H.
In the random phase approximation,  we obtain
$$
F_{\alpha\beta}(q)=
P_\alpha(q)\delta_{\alpha\beta}+\sum_\gamma
P_\alpha(q) V_{\alpha \gamma }({\bm q})F_{\gamma \beta},
\eqno(4.2)
$$
$$
P_\alpha(q)
\cong {\beta(\omega^0_{\alpha{\bm q}})^2 \over 
q_0(q_0+\frac{\rm i}{\tau_{\alpha0}})-(\omega^0_{\alpha{\bm q}})^2},
\eqno(4.3)
$$
\begin{align}
V_{\alpha\beta}({\bm q}) &=<\frac1{\sqrt{N_\alpha N_\beta}}
\sum_{m\neq n}
e^{-{\rm i}{\bm q}\cdot ({\bm R}_m-{\bm R}_n)}
V_{\alpha \beta}({\bm R}_m-{\bm R}_n)>_c
\notag \\
&=\sqrt{N_\alpha \over N_\beta}
\int d^3Rg_{\alpha \beta}(R)
e^{{\rm i}{\bm q}\cdot 
{\bm R} } V_{\alpha \beta} ({\bm R})
\tag{4.4}
\end{align}
$$
V_{\alpha \beta}({\bm R}_{\alpha m}-{\bm R}_{\beta n})
=\int d^3xd^3y \tilde w_{\alpha m}^2(\bm{x}-\bm{R}_{\alpha m})
V_{\alpha\beta}(\bm{x}-\bm{y})
\tilde w^2_{\beta n}(\bm{y}-\bm{R}_{\beta n}),
\eqno(4.5)
$$
$$
\frac1{2\tau_{\alpha 0} }=
\sqrt{N_\alpha (U_\omega^{\alpha 00}+U_J^{\alpha 00})}
\eqno(4.6)
$$
$$
(\omega_{\alpha q}^0)^2 \cong 
{(v_{\alpha {\rm p}} )^2 \over 3} q^2,
\qquad v_{\alpha {\rm p}} =\frac1{N}\sum_{\bm p}
{\partial \epsilon_{\alpha 0{\bm p}} \over \partial p},
\eqno(4.7)
$$
where we have analytically continued
${\rm i}\nu_n \rightarrow q_0$ in Eq.(4.3).
$\tau_{\alpha 0}$ and $v_{\alpha {\rm p}}$ are the relaxation
time and the mean
velocity of the $\alpha$-particle. Eq.(4.3) is presented in
\cite{K1}.
Hereafter, we omit the suffices
$d$ and $0$ in $V^d_{\alpha 0 \beta 0}$.

Appling Eq.(4.2) to water, we obtain the dynamical equation:
$$
\begin{pmatrix}
1- P_{\rm O}V_{\rm OO} & 
P_{\rm O} V_{\rm OH} \\
P_{\rm H} V_{\rm HO} 
& 1- P_{\rm H} V_{\rm HH}
\end{pmatrix}
\begin{pmatrix}
F_{\rm OO} & F_{\rm OH} \\
F_{\rm HO} & F_{\rm HH}
\end{pmatrix}=
\begin{pmatrix}
P_{\rm O} & 0 \\
0 & P_{\rm H} 
\end{pmatrix}
\eqno(4.8)
$$
with the solution:
$$
\begin{pmatrix}
F_{\rm OO} & F_{\rm OH} \\
F_{\rm HO} & F_{\rm HH}
\end{pmatrix}
=\frac1{\det |\quad |}
\begin{pmatrix}
(1- P_{\rm H} V_{\rm HH})P_{\rm O} & 
P_{\rm O} V_{\rm OH} P_{\rm H} \\
P_{\rm H} V_{\rm HO} P_{\rm H}
& (1- P_{\rm O}V_{\rm OO})P_{\rm H}
\end{pmatrix},
\eqno(4.9)
$$
$$
\det |\quad|=(1-P_{\rm O} V_{\rm OO})
(1- P_{\rm H} V_{\rm HH})
- P_{\rm O} V_{\rm OH}P_{\rm H} V_{\rm HO}.
\eqno(4.10)
$$
Here in order to construct the hydrogen bonding network,
we should consider that $V_{\rm OH}(R)$ in Eq.(4.4) consists of
hydrogen covalent bonding, $V_{\rm OH}^{\rm hcb}(R)$ and
hydrogen bonding, $V_{\rm OH}^{\rm hb}(R)$, the pair distribution
functions of which are  $g_{\rm OH}^{\rm hbc}(R)$
and $g_{\rm OH}^{\rm hc}(R)$, respectively.  Corresponding
to this fact, we consider that $V_{\rm OH}({ q})$ consists of
hydrogen covalent bonding  $V_{\rm OH}^{\rm hcb}({ q})$ and
hydrogen bonding $V_{\rm OH}^{\rm hb}({ q})$. Thus we put
$$
V_{\rm OH}({ q})=V_{\rm OH}^{\rm hcb}({ q}), \ 
V_{\rm OH}^{\rm hb}({ q}).
\eqno(4.11)
$$
But in deriving Eq.(4.8),
we have used  Eq.(4.4) under the nearest neighbour
approximation so that the terms in Eqs.(4.9, 10)
involves unphysical cross terms
such as $V_{\rm OH}^{\rm hcb}V_{\rm HO}^{\rm hcb}$ and
$V_{\rm OH}^{\rm hb}V_{\rm HO}^{\rm hb}$.
When we calculate Eq.(4.1),
we should configurationally average all possible
Feynmann diagrams
over the position of particles ${\bm R}_{\alpha m}$. 
If we configurationally average the term
$<V_{\rm OH}V_{\rm HO}>$ 
over ${\bm R}_{{\rm H}m}$ considering
the pair distribution functions $g_{\rm OH}^{\rm hbc}(R)$
and $g_{\rm OH}^{\rm hc}(R)$, the most dominant terms in
$<V_{\rm OH}V_{\rm HO}>_c$ come from
the terms $V_{\rm OH}^{\rm hbc}(a_{\rm OH}^{\rm hcb})
V_{\rm HO}^{\rm hc}(a_{\rm HO}^{\rm hc})$ and
$V_{\rm OH}^{\rm hc}(a_{\rm OH}^{\rm hb})
V_{\rm HO}^{\rm hbc}(a_{\rm HO}^{\rm hbc})$,
where $a^{\rm hcb, hb}_{\rm OH}$
are the mean distances of hydrogen covalent and
hydrogen bond, respectively. Since in our calculation
we concentrate ourselves to the hydrogen bonding network,
we should take into account of both the terms  
$V_{\rm OH}^{\rm hbc}V_{\rm HO}^{\rm hc}$ and
$V_{\rm OH}^{\rm hc}V_{\rm HO}^{\rm hbc}$ as pairs.
Thus when the term $V_{\rm OH}V_{\rm HO}$ appears, 
$V_{\rm OH}V_{\rm HO}$ should be replaced by
$V_{\rm OH}^{\rm hbc}V_{\rm HO}^{\rm hc}+
V_{\rm OH}^{\rm hc}V_{\rm HO}^{\rm hbc}$.

We can obtain the modes by solving
the secular equation of Eq. (4.8):
$$
\{1-P_{\rm O}(q)V_{\rm OO}({ q})\}
\{1-P_{\rm H}(q)V_{\rm HH}({ q})\}-
P_{\rm O}(q)P_{\rm H}(q) V_{\rm OH}({q})V_{\rm HO}({q})=0.
\eqno(4.12)
$$

Before we solve Eq.(4.12), we investigate the secular equation for
the sound in a system consist of only $\alpha$-type of particles:
$$
1+P_\alpha(q)V_{\alpha\alpha}({q})=0
\eqno(4.13)
$$
We investigate Eq.(4.13) in the two limiting cases:

(i) $q_0\tau_{\alpha 0} \gg 1$

Eq.(4.13) leads to the sound frequency for $\alpha$-particles:
$$
q^2=\omega_{\alpha {\rm s}}^2(q) =
(1+\beta V_{\alpha\alpha}({ q}))(\omega^0_{\alpha q})^2
={(\omega^0_{\alpha q})^2 \over S_\alpha(q)} =
v^2_{\alpha {\rm s}}q^2,
\eqno(4.14)
$$
where $S_\alpha(q)$ and $v_{\alpha {\rm s}}$
are the static structure factor and
the sound velocity of the $\alpha$-particle:
$$
S_\alpha(q)\equiv (1+\beta V_{\alpha\alpha}({ q}))^{-1},   
\eqno(4.15)
$$
$$
v_{\alpha {\rm s}}={v_{\alpha {\rm p}} \over \sqrt{3S_\alpha}(0)}.
\eqno(4.16)
$$

Next we investigate the sound instability.
In the first aproximation, if we consider a first
peak for a pair distribution function, we can put
$$
g_{\alpha \alpha}(R)=
\rho_\alpha\delta(R-a_{\alpha \alpha}),
\eqno(4.17)
$$
where $a_{\alpha\alpha}$ is the mean distance
between the $\alpha$- particles.
Then we obtain
$$
V_{\alpha \alpha}(q)=4 \pi a_{\alpha \alpha}^2\rho_\alpha
V_{\alpha \alpha}(a_{\alpha \alpha})
{\sin a_{\alpha \alpha}q \over a_{\alpha \alpha}q},
\eqno(4.18)
$$
where $V_{\alpha \alpha}$ has the minimum negative value
at a reciprocal particle distance
$\tilde K_{\alpha \alpha} \cong 3\pi/2a_{\alpha\alpha}$.
Since 
$$
\lim_{q_0\rightarrow 0} P_\alpha(0) \cong -\beta,
\eqno(4.19)
$$
and $V_{\alpha \alpha}$ has the minimum negative value
at a reciprocal particle distance
$\tilde K_{\alpha \alpha}$,
the sound instability in the $\alpha$-liquid occurs at the temperature
$T_{\alpha 0}$ and
the reciprocal particle distance
$\tilde K_{\alpha \alpha}$ in Eq.(4.15):
$$
1+\beta_{\alpha 0}V_{\alpha \alpha}(\tilde K_{\alpha \alpha})=0.
\eqno(4.20)
$$

(ii) $q_0\tau_{\alpha 0} \ll 1$

Eq.(4.13) leads to the diffusion mode
$$
q_0={\rm i}\tau_{\alpha 0}v_{\alpha {\rm s}}^2 q^2,
\eqno(4.21)
$$
where the  diffusion coefficient is given by
$$
D_\alpha =\tau_{\alpha 0}v_{\alpha {\rm s}}^2.
\eqno(4.22)
$$

Next we investigate the secular equation Eq.(4.12)
for water in the two limiting cases:

(I) $q_0 \tau_{\alpha 0} \gg 1$

We obtain the sound modes:
\begin{align}
q^2_0 &= \omega_{{\rm s}\pm} ^2(q)        \notag \\
&=\frac12\left[\omega_{\rm O}^2(q)+\omega_{\rm H}^2(q)
\pm\sqrt{(\omega_{\rm O}^2(q)-
\omega_{\rm H}^2(q))^2+4\beta^2
(\omega^0_{{\rm O}q})^2(\omega^0_{{\rm H}q})^2
 V_{\rm OH}({ q})V_{\rm HO}({ q})} \right]     \notag \\
&=v_{{\rm s}\pm}^2 q^2,  
\tag{4.23}
\end{align}
where $v_{{\rm s}\pm}$ are the sound velocities:
$$
v_{{\rm s}\pm}=\sqrt{\frac12\{v_{{\rm O}{\rm s}}^2+
v_{{\rm H}{\rm s}}^2
\pm\sqrt{(v_{{\rm O}{\rm s}}^2-v_{{\rm H}{\rm s}}^2)^2+
\frac49\beta^2
v_{{\rm Op}}^2v_{{\rm Hp}}^2V_{\rm OH}({ q})
V_{\rm HO}({ q})}\} }.
\eqno(4.24)
$$
If the term $V_{\rm OH}V_{\rm HO}$ dominates,
$\omega_{\rm s-}$ and $v_{\rm s-}$ disappear.
Next we investigate the sound instability.
In the first aproximation, if we consider a
peak for the pair distribution function, we can put
$$
g_{\alpha \beta}(R)=\rho_\beta\delta(R-a_{\alpha \beta}),
\eqno(4.25)
$$
where $a_{\alpha\beta}$ is the mean distance between
the $\alpha$ and $\beta$ particles.
Then we obtain
$$
V_{\alpha \beta}(q)=4 \pi a_{\alpha \beta}^2\rho_\beta
V_{\alpha \beta}(a_{\alpha \beta})
{\sin a_{\alpha \beta}q \over a_{\alpha \beta}q},
\eqno(4.26)
$$
where $V_{\alpha \beta}$ has the minimum negative value
at a reciprocal particle distance
$\tilde K_{\alpha \beta} \cong 3\pi/2a_{\alpha\beta}$.

In water the interaction potential $V^{\rm hb}_{\rm OH}$
and the pair distribution function $g^{\rm hb}_{\rm OH}$ in
$V_{\rm OH}$ are the most important. 
The peak of $g^{\rm hb}_{\rm OH}$ corrsponds to the hydrogen bond,
which constitute the cluster of hydrogen bonded molecules.
If we denote  the
mean distance corresponding to the peak  by
 $a^{\rm hb}_{\rm OH}$, the reciprocal particle distance
is given by $\tilde K^{\rm hb}_{\rm OH}$.
The sound instability occurs when the secular
equation Eq.(4.12) with
Eq.(4.19) becomes zero at a 
temperature $T_0$ and at a wave vector 
$\tilde K\cong \tilde K_{\rm OH}^{\rm hb}$:
$$
\{1+\beta_0 V_{\rm OO}(\tilde K)\}
\{1+\beta_0 V_{\rm HH}(\tilde K) \}-
\beta^2_0  V_{\rm OH}(\tilde K)V_{\rm HO}(\tilde K)=0.
\eqno(4.27)
$$
$T_0$ is given by
$$
T_0=\frac1{2k_{\rm B}}\{V_{\rm OO}(\tilde K)+
V_{\rm HH}(\tilde K)+
\sqrt{(V_{\rm OO}(\tilde K)-V_{\rm HH}(\tilde K))^2+
4V_{\rm OH}(\tilde K)V_{\rm HO}(\tilde K)} \}.
\eqno(4.28)
$$

(II) $q_0 \tau_{\alpha 0} \ll 1$

We obtain the diffusion modes:
\begin{align}
q_0  =    {\rm i}D^\pm q^2 =
 \frac{\rm i}2\{&\tau_{{\rm O}0} \omega_{\rm O}^2(q) 
        +\tau_{\rm H0} \omega_{\rm H}^2(q)   \notag \\
      &\pm \sqrt{(\tau_{\rm O0}\omega_{\rm O}^2(q)
-\tau_{\rm {H0}}
 \omega_{\rm H}^2(q))^2+4\beta^2
\tau_{\rm O0}(\omega^0_{{\rm O}q})^2
\tau_{\rm H0}(\omega^0_{{\rm H}q})^2
 V_{\rm OH}({ q})V_{\rm HO}({ q})} \}q^2,
 \tag{4.29}
\end{align}
where $D^\pm$ are diffusion coefficients:
\begin{align}
D^{\pm}&=\frac12\{\tau_{\rm O0}v_{{\rm Os}}^2+
\tau_{\rm H0}v_{{\rm Hs}}^2\pm
\sqrt{(\tau_{\rm O0}v_{{\rm Os}}^2-
\tau_{\rm H0}v_{{\rm Hs}}^2)^2+\frac49\beta^2
\tau_{\rm O0}v_{{\rm Op}}^2\tau_{\rm H0} v_{{\rm Hp}}^2
V_{\rm OH}({ q})V_{\rm HO}({ q})} \}
\notag \\
&=\frac12\{D_{\rm O}+D_{\rm H} 
\pm \sqrt{(D_{\rm O}-D_{\rm H})^2+\frac49\beta^2
\tau_{\rm O0}v_{{\rm Op}}^2\tau_{{\rm H}0} v_{{\rm Hp}}^2
V_{\rm OH}({ q})V_{\rm HO}({ q})} \}.
\tag{4.30}
\end{align}
If the term $V_{\rm OH}V_{\rm HO}$ dominates, $D^-$ disappears.

\section{Phonons, boson peaks and viscosity}
\hspace{5ex}
Next we investigate the phonon Green's functions, which are defined by
$$
D_{\alpha i\beta j{\bm q}}(\tau_1-\tau_2)\equiv
 -\frac1{\hbar \sqrt{
N_{\alpha}N_\beta}}<T_\tau\rho_{\alpha i{\bm q}}
(\tau_1)\rho^\dagger_{\beta j{\bm q}}(\tau_2)>_c \equiv
\frac1{\beta\hbar}
\sum_{{\rm i}\nu_n}e^{-{\rm i}\nu_n(\tau_1-\tau_2)}
D_{\alpha i\beta j}(q),
\eqno(5.1)
$$
In the random phase approximation, we obtain
 the phonon Green's functions and the gap
equation [1]:
$$
D_{\alpha i\beta j}(q)= Q_\alpha(q)\delta_{\alpha\beta}\delta_{ij}
+ Q_\alpha(q)\sum_{\gamma l}
<V_{\alpha i\gamma l}^{od}({\bm q})>_cD_{\gamma l\beta j}(q),
\eqno(5.2)
$$
$$
{\tilde \Delta_\alpha \over \sqrt{N_\alpha}}\delta_{ij}= 
Q_\alpha(0)<V_{\alpha i\beta j}^{od}(0)>_c
{\tilde \Delta_\beta \over \sqrt{N_\beta}}, 
\eqno(5.3)
$$
where 
$$
V_{\alpha i\beta j}^{od}({\bm q}) =
<\frac{\zeta_\alpha \zeta_\beta}
{\sqrt{N_\alpha N_\beta}} \sum_{m\neq n}
e^{-{\rm i}{\bm q}\cdot ({\bm R}_m-{\bm R}_n)}
V_{\alpha i\beta j}({\bm R}_m-{\bm R}_n)>_c,
\eqno(5.4)
$$
$$
V_{\alpha i\beta j}({\bm R}_{\alpha m}-{\bm R}_{\beta n})
\equiv\int\,d^3xd^3y\nabla_i
\tilde w^2_{\alpha m0}({\bm x}-{\bm R}_{\alpha m})V_{\alpha\beta}(\bm{ x}-\bm{y})
\nabla_j \tilde w^2_{\beta 0}({\bm y}-{\bm R}_{\beta n}).
\eqno(5.5)
$$
$$
Q_\alpha(q)={2\omega_\alpha/\hbar 
\over (q_0+{\rm i}/2\tau_{\alpha {\rm M}})^2
-\omega_\alpha^2},
\eqno(5.6)
$$
where $\tau_{\alpha {\rm M}}$ is called the Maxwell relaxation time and
$$
\frac1{2\tau_{\alpha {\rm M}}}\equiv 
\frac1{2\tau_{\alpha 0}}+\frac1{2\tau_{\alpha 1}}
-\frac1{2\tau_{\alpha 0}^o}-\frac1{2\tau_{\alpha 1}^o}.
\eqno(5.7)
$$
$\tau_{\alpha 0,1}$ are the relaxation times of the particles
in the lower or upper bands,
respectively. $\tau_{\alpha 0,1}^o$ comes from the simultaneously
scattering processes of particles between the lower and upper
bands. This term corresponds to the cosine-function term
with the scattering angle of an electron in the electric conductivity.

At low temperatures, since 
$\tilde \Delta_\alpha={N_\alpha \over \zeta_\alpha}$, 
the gap equation (5.3) leads to
$$
1= \zeta^2_\alpha Q_\alpha(0)\sum_{\beta n}
< V_{\alpha i\beta i}({\bm R}_{\beta n})>_c.
\eqno(5.8)
$$
Substituting $\zeta_\alpha =<\zeta_{\alpha n}>_c$ in 
 Eqs.(2.6) and (5.6) into Eq.(5.8) under 
$\omega_\alpha \tau_{\alpha {\rm M}} \ll 1$,
we obtain
$$
M_\alpha\omega^2_\alpha=-\sum_{\beta n}<V_{\alpha i \beta i}
({\bm R}_{\beta n})>_c.
\eqno(5.9)
$$
Eq.(5.9) means that every 
$\alpha $-particle feels the same statistical averaged potential.
From Eq.(5.4, 5) the matrix element of the dynamical equation (5.2) is
written  as
$$
\delta_{\alpha\beta}\delta_{ij}- Q_\alpha(q)
<V^{od}_{\alpha i\beta j}({\bm q})>_c=
-{\zeta_\alpha^2}Q_\alpha(q)M_\alpha[q_0(q_0+
\frac{\rm i}{\tau_{\alpha {\rm M}}})
\delta_{\alpha\beta}\delta_{ij} -M_{\alpha i\beta j}({\bm q})],
\eqno(5.10)
$$
where $M_{\alpha i\beta j}$ is the dynamical matrix for phonons. 
At low temperatures the dynamical matrix is given by
$$
M_{\alpha i\beta j}({\bm q})={1 \over M_\alpha}\sum_\gamma
\int\,d^3Rg_{\alpha
\gamma}(R)\{\delta_{\beta\gamma}
{\zeta_\gamma \over \zeta_\alpha}
\sqrt{N_\alpha \over N_\gamma}
e^{{\rm i}\bm{ q\cdot R}}-
\delta_{\alpha\beta}\delta_{ij}\}
V_{\alpha i\gamma j}({\bm R}).
\eqno(5.11)
$$

Here, as discussed on the term $V_{\rm OH}({ q})$ in Eq.(4.11)
in taking into account of the hydrogen bonding network, we
should consider that the term  $M_{{\rm O}i{\rm H}j}({\bm q})$ consists
of  the term $M_{{\rm O}i{\rm H}j}^{hcb}({\bm q})$ corresponding to
the hydrogen covalent bonding and the term
$M_{{\rm O}i{\rm H}j}^{hb}({\bm q})$ to the hydrogen bonding.
Hereafter, in the same manner as the sound,
if the cross term 
$M_{{\rm O}i{\rm H}j}({\bm q})M_{{\rm H}i{\rm O}j}({\bm q})$
appears,
$M_{{\rm O}i{\rm H}j}({\bm q})M_{{\rm H}i{\rm O}j}({\bm q})$ 
should be replaced by
$M_{{\rm O}i{\rm H}j}^{\rm hcb}({\bm q})
M_{{\rm H}i{\rm O}j}^{\rm hb}({\bm q})+
M_{{\rm O}i{\rm H}j}^{\rm hb}({\bm q})
M_{{\rm H}i{\rm O}j}^{\rm hcb}({\bm q})$.

Now we investigate Eq.(5.5). We separate the potential $V$ into
the short range one, $V^N$ and the Coulombic one, $V^C$:
$$
V_{\alpha i\beta j}({\bm R}_{\alpha m}-{\bm R}_{\beta n})
=V_{\alpha i\beta j}^N ({\bm R}_{\alpha m}-{\bm R}_{\beta n})
+V_{\alpha i\beta j}^C({\bm R}_{\alpha m}-{\bm R}_{\beta n}).
\eqno(5.12)
$$
When we take into account of the polarization
of the hydogen covalent bonds, we make the rigid ion approximation,
the Ewald's method \cite{K5,Mar,Ewa}.
In the rigid ion approximation;
$w^2_{\alpha 0}({\bm R})
\longrightarrow \delta(\bm{R})$, we obtain
$$
V_{\alpha i \beta j}^C({\bm R})=-{\partial^2 
\over \partial x_i \partial x_j}
{e_\alpha e_\beta \over 4\pi \epsilon_0 R},
\eqno(5.13)
$$
where we have used the SI units, $\epsilon_0$
is the dielectric constant
in vacuum and $e_\alpha$ is the charge of
an $\alpha$-particle; $e_{\rm O}=-2e_{\rm H}$.
We define the Fourier transformed potential as
$$
\hat V_{\alpha i \beta j}({\bm q})=<\sum_n 
e^{-{\rm i}{\bm q}\cdot
({\bm R}_{\alpha m}-{\bm R}_{\beta n})}V_{\alpha i \beta j}
({\bm R}_{\alpha m}-{\bm R}_{\beta n})>_c
=\hat V_{\alpha i \beta j}^N({\bm q})+
\hat V_{\alpha i \beta j}^C({\bm q}).
\eqno(5.14)
$$

Next we investigate the Coulomb potential:
$$
\varphi^C_{\alpha \beta}(R)={e_\alpha e_\beta 
\over 4\pi \epsilon_0 R},
\qquad \hat \varphi_{\alpha \beta}^C(q) ={e_\alpha e_\beta
 \over \epsilon_0 q^2},
\eqno(5.15).
$$
which satisfy 
$$
\varphi^C_{\alpha \beta}(R)=\frac1{(2\pi)^3}\int d^3q
e^{{\rm i}\bm {q\cdot R}}\hat \varphi^C ( q).
\eqno(5.16)
$$
In the Ewald method, we define a potential,
$\varphi^G_{\alpha \beta}(R)$,
which is the potential energy of a point charge
$e_\alpha$ at a distance $R$
from a three dimentional Gaussian distribution
$\rho^G_\beta(R)$ with the
total charge $e_\beta$:
$$
\rho^G_\beta(R)=e_\beta(\frac{P}{\pi})^{3/2}e^{-PR^2},\qquad 
\hat \rho^G_\beta(q)=e_\beta e^{-q^2/4P},
\eqno(5.17)
$$
which satisfy
$$
\rho^G_\beta(R)=\frac1{(2\pi)^3}\int d^3 q\hat \rho^G_\beta(q)
e^{{\rm i} \bm{q\cdot R}}.
\eqno(5.18)
$$
Thus we obtain
$$
\varphi^G_{\alpha \beta}(R)={e_\beta \over 4\pi\epsilon_0}[
\frac1R\int^R_0\rho^G_\beta(R')d^3R'+\int^\infty_R
{\rho^G_\beta(R') \over R'}d^3R']. 
\eqno(5.19)
$$
The Poisson equation:
$$
\nabla^2 \varphi^G_{\alpha \beta}(R)=
\frac1{(2\pi)^3}\int d^3 q(-q^2
\hat \varphi^G_{\alpha \beta}(q))e^{{\rm i} \bm{q \cdot R}}
=-{e_\alpha \over \epsilon_0}\rho^G_\beta (R),
\eqno(5.20)
$$
and Eq.(5.17) lead to
$$
\hat\varphi^G_{\alpha\beta}(q)={e_\alpha e_\beta \over 
\epsilon_0 q^2}e^{-q^2/4P}.
\eqno(5.21)
$$
Now we introduce
$$
\varphi^C_{\alpha\beta}(R)=\varphi^G_{\alpha\beta}(R)+
\varphi^C_{\alpha\beta}(R)-\varphi^G_{\alpha\beta}(R)
\equiv \varphi^G_{\alpha\beta}(R)+
\varphi^H_{\alpha\beta}(R)
\eqno(5.22)
$$
Using Eqs.(5.15), (5.19) and 
$\int_0^\infty d^3R'\rho_\beta^G(R')
=e_\beta$, we obtain
$$
\varphi^H_{\alpha\beta}(R)={e_\alpha e_\beta \over
2\epsilon_0 \pi^{3/2}} \frac1R\int^\infty_{\sqrt{P}R}
e^{-s^2}ds.
\eqno(5.23)
$$
Here we define
\begin{align}
\hat V_{\alpha i\beta j}^C({\bm q}) &=-<\sum_n
e^{-{\rm i}{\bm q}\cdot ({\bm R}_{\alpha m}
-{\bm R}_{\beta_n})}
{\partial^2 \over \partial x_i \partial x_j}
\{\varphi^G_{\alpha \beta}(R)+
\varphi^H_{\alpha \beta}(R)\}|_{{\bm R}
={\bm R}_{\alpha m}-{\bm R}_{\beta_n}} >_c  \notag \\
&\equiv \hat V_{\alpha i\beta j}^G({\bm q})+
\hat V_{\alpha i\beta j}^H({\bm q}).
\tag{5.24}
\end{align}
First we  investigate the term $\hat V^C$, which is written as
\begin{align}
\hat V_{\alpha i\beta j}^G({\bm q}) &=-<\sum_n
e^{-{\rm i}{\bm q}\cdot ({\bm R}_{\alpha m}
-{\bm R}_{\beta_n})}
{\partial^2 \over  \partial x_i \partial x_j}
\{\frac1{(2\pi)^3}\int d^3q'
e^{{\rm i}\bm {q'\cdot R}}\hat \varphi^G ({\bm q'})
\}|_{{\bm R}={\bm R}_{\alpha m}-
{\bm R}_{\beta_n}} >_c \notag \\
& =<\sum_n e^{{\rm i}\bm{ (q'-q)}
\cdot ({\bm R}_{\alpha m}-{\bm R}_{\beta_n})}
{e_\alpha e_\beta \over (2\pi)^3 \epsilon_0}
\int d^3q'{q'_i q'_j \over q'^2}e^{-q'^2/4P}>_c \notag \\
& =<\sum_n \int d^3q'[ {q'_i q'_j 
\over q'^2}+{q'_i q'_j \over q'^2}\{e^{-q'^2/4P}-1\}]
e^{{\rm i}\bm{(q'-q)}\cdot ({\bm R}_{\alpha m}-{\bm R}_{\beta_n})}>_c
\notag \\
& ={e_\alpha e_\beta \over (2\pi)^3 \epsilon_0 v_a}\int d^3R
g_{\alpha \beta}(R) \int d^3q'[ {q'_i q'_j 
\over q'^2}+{q'_i q'_j \over q'^2}\{e^{-q'^2/4P}-1\}]
e^{{\rm i}\bm{(q'-q)}\cdot {\bm R}}
\notag \\
&\cong {e_\alpha e_\beta \rho_\beta \over \epsilon_0 v_a}[
{q_i q_j \over q^2}+{q_i q_j \over q^2}\{e^{-q^2/4P}-1\}],
\tag{5.25}
\end{align}
where we have made the approximation;
$g_{\alpha \beta}(R)\cong \rho_\beta$
and $v_a$ is a primitive unit cell.
The term $V^H$ is given by
\begin{align}
V^H_{\alpha i\beta j}({\bm q}) &=-<\sum_n
e^{-{\rm i}{\bm q}\cdot 
({\bm R}_{\alpha m}-{\bm R}_{\beta_n})}
{\partial^2 \over  \partial x_i \partial x_j}
\{\frac1{(2\pi)^3}\int d^3q'
e^{{\rm i}\bm {q'\cdot R}}\hat \varphi^H ({\bm q'})
\}|_{{\bm R}={\bm R}_{\alpha m}-{\bm R}_{\beta_n}}>_c
\notag \\
& =<\sum_n e^{-{\rm i}\bm{ (q)}\cdot 
({\bm R}_{\alpha m}-{\bm R}_{\beta_n})}
{\partial^2 \over  \partial x_i \partial x_j}
{e_\alpha e_\beta \over (2\pi)^3 \epsilon_0}
\frac1R\int_{\sqrt{P}R}^\infty ds e^{-s^2}|_{{\bm R}=
{\bm R}_{\alpha m}-{\bm R}_{\beta_n}}>_c
\notag \\
& ={e_\alpha e_\beta \over 
(2\pi)^3 \epsilon_0 v_a}P^{3/2}\int d^3R
g_{\alpha \beta}(R) H_{ij}(\sqrt{P} R)
e^{-{\rm i}\bm{q}\cdot {\bm R}}
\notag \\
& \cong {e_\alpha e_\beta\rho_\beta \over (2\pi)^3
\epsilon_0 v_a}P^{3/2}\int d^3R
 H_{ij}(\sqrt{P} R)e^{-{\rm i}\bm{q}\cdot {\bm R}},
\tag{5.26}
\end{align}
$$
H_{ij}(x)\equiv {\partial^2 \over \partial x_i \partial x_j}
\frac2{\sqrt{\pi}}\frac1{x}
\int_x^\infty ds e^{-s^2}.
\eqno(5.27)
$$
It should be noted the function $H_{ij}(x)$
is singular at $x=0$. When we treat
the crystalline state \cite{Mar,Ewa},
$H_{ij}(x)$ is
replaced by $H^0_{ij}=
{\partial^2 \over \partial x_i \partial x_j}
\frac2{\sqrt{\pi}}\frac1{x}
\int_0^x ds e^{-s^2}$, because we should exclude the singularity, since
the dynamical matrix includes ${\bm R}_m={\bm R}_n$.
But in our case, taking into account of the spherical symmetry
$\sum_i H_{ii}=\frac4{\sqrt{\pi}}e^{-x^2}$ and the property of the argument
of integration $d^3R=4\pi R^2 dR$ yields the integration finite.
Thus Eqs.(5.25) and (5.26) lead to
$$
V^C_{\alpha i\beta j}(q)=e_\alpha e_\beta 
\rho_\beta [{q_i q_j \over 
v_a \epsilon_0 q^2}-Q_{ij}(q)],
\eqno(5.28)
$$
$$
Q_{ij}(q)=-{q_i q_j \over \epsilon_0 v_a}
[e^{-{q^2 \over 4P}} -1]+
{P^{3/2} \over 4\pi \epsilon_0 v_a}\int d^3 R
H_{ij}(\sqrt{P}R)e^{-{\rm i}\bm{q\cdot R}},
\eqno(5.29)
$$
where the first term and the second term
in Eq.(5.28) relate to the macroscopic
electric field and the Lorentz field, respectively.
In the spherical symmetry, we obtain
$$
Q_{ij}(q)=\delta_{ij} Q(q) \cong 
\delta_{ij}\frac1{3\epsilon_0 v_a}.
\eqno(5.30)
$$
There is the electrical charge neutrality:
$$
\sum_\alpha e_{\alpha}\rho_{\alpha}=0.
\eqno(5.31)
$$

Next we investigate the short range potential $V^N$. In order to do so,
we introduce the couplings like the chemical bondings:
$$
V_{\sigma\alpha\beta}(R)=-\int\, d^3xd^3y\nabla_{x_3}
\tilde w_{\alpha0}^2({\bm x})V_{\alpha\beta}(\bm{ x}-\bm{y})
\nabla_{y_3}\tilde w_{\beta0}^2({\bm y}-R{\bm e}_3),
\eqno(5.32)
$$
$$
V_{\pi\alpha\beta}(R)=-\int\,d^3xd^3y\nabla_{x_1}
\tilde w_{\alpha0}^2({\bm x})V_{\alpha\beta}(\bm{ x}-\bm{y})
\nabla_{y_1}\tilde w_{\beta0}^2({\bm y}-R{\bm e}_3).
\eqno(5.33)
$$
Here we take ${\bm q}=q{\bm e}_z$ and $\xi=qR$.
Since we are concerned with the spherical symmetry distribution of
particles, we obtain
\begin{align}
M_{\alpha x\alpha x}({\bm q}) &=M_{\alpha y\alpha y}({\bm q})
=-\frac1{M_\alpha} \int\,d^3Rg_{\alpha\alpha}(R)
\left[ V_{\sigma\alpha\alpha}(R) 
\{-{\cos \xi \over \xi^2}
+{\sin\xi \over \xi^3}\} \right.  \notag \\
&  +\left. V_{\pi\alpha\alpha}(R)
\{{\sin\xi \over \xi}+{\cos\xi \over \xi^2}-
{\sin\xi \over \xi^3}\} +e^2_\alpha 
\rho_\alpha \frac1{3\epsilon_0 v_a}
\right] \notag \\
& +\frac1{M_\alpha}\sum_\gamma 
\int d^3Rg_{\alpha\gamma}(R) \{
{V_{\sigma\alpha\gamma}(R) \over 3}+
{2V_{\pi\alpha\gamma}(R) \over 3} \},
\tag{5.34}
\end{align}
\begin{align}
M_{\alpha z\alpha z}({\bm q}) &=-\frac1{M_\alpha}\int\,d^3R
g_{\alpha\alpha}(R)\left[V_{\sigma\alpha\alpha}(R)
\{{\sin\xi \over \xi}+{2\cos\xi \over \xi^2}-{2\sin\xi \over
\xi^3}\}\right.  \notag \\
& + \left.V_{\pi\alpha\alpha}(R) 
\{ -{2\cos\xi \over \xi^2}+{2\sin\xi \over \xi^3} \} 
-e^2_\alpha \rho_\alpha \frac2{3\epsilon_0 v_a} \right] \notag  \\
& +\frac1{M_\alpha}\sum_\gamma\int d^3Rg_{\alpha\gamma}(R)\{
{V_{\sigma\alpha\gamma}(R)
\over 3}+{2V_{\pi\alpha\gamma}(R) \over 3} \},
\tag{5.35}
\end{align}
\begin{align}
M_{\alpha x\beta x}({\bm q}) &=M_{\alpha y\beta y}({\bm q})=
-\frac1{M_\alpha}{\zeta_\beta \over \zeta_\alpha}
\sqrt{N_\alpha \over
N_\beta}\int\,d^3Rg_{\alpha\beta}(R) \notag \\
& \times \left\{V_{\sigma\alpha\beta}(R)
(-{\cos\xi \over \xi^2}+{\sin\xi \over \xi^3})
+V_{\pi\alpha\beta}(R)({\sin\xi \over \xi}+{\cos\xi \over 
\xi^2}-{\sin\xi \over \xi^3}) +
e_\alpha e_\beta \rho_\beta \frac2{3\epsilon_0 v_a} \right\},
\tag{5.36}
\end{align}
\begin{align}
M_{\alpha z\beta z}({\bm q}) &=-\frac1{M_\alpha}
{\zeta_\beta \over \zeta_\alpha}\sqrt{N_\alpha \over
N_\beta}\int\,d^3Rg_{\alpha\beta}(R)    
  \left\{V_{\sigma\alpha\beta}(R)
({\sin\xi \over \xi}+{2\cos\xi \over \xi^2}
-{2\sin\xi \over \xi^3}) \right. \notag \\
& +\left.V_{\pi\alpha\beta}(R)(-{2\cos\xi \over 
\xi^2}+{2\sin\xi \over \xi^3}) 
-e_\alpha e_\beta \rho_\beta \frac2{3\epsilon_0 v_a} \right\},
\tag{5.37}
\end{align}
where the other terms are zero. Since we have taken
${\bm q}=q{\bm e}_3$, the z-component of the dynamical
matrix corresponds to the longitudinal mode and x and y-components
to the transverse modes. It should be noted that the dynamical matrices have
the effects of the electric polarization in the last term in the curly brackets.
But because of the charge neutrality (5.31),
the last terms in Eqs.(5.34) and (5.35) do not have the effect of the polarization.

The secular equation of Eq.(5.2) leads to
$$
\{q_0(q_0+\frac{\rm i}{\tau_{{\rm OM}}})-M_{{\rm O}i{\rm O}i}({\bm q})\}
\{q_0(q_0+\frac{\rm i}{\tau_{{\rm HM}}})-M_{{\rm H}i{\rm H}i}({\bm q})\}
-M_{{\rm O}i{\rm H}i}({\bm q})M_{{\rm H}i{\rm O}i}({\bm q}) =0,
\eqno(5.38)
$$
where $i=z$ corresponds to a longitudinal mode and $i=x,\ y$
to two transverse modes. Now we investigate the secular equation (5.38)
in the two limiting cases:

(I) $q_0\tau_{\alpha {\rm M}} \gg 1$:

We obtain phonon modes
\begin{align}
q^2_0 &=\omega_{i\pm}^2({\bm q}) \notag \\
             &=\frac12\left[M_{{\rm O}i{\rm O}i}
({\bm q})+M_{{\rm H}i{\rm H}i}({\bm q})\pm\sqrt{
\{ M_{{\rm O}i{\rm O}i}({\bm q})-M_{{\rm H}i{\rm H}i}({\bm q}) \}^2+
4M_{{\rm O}i{\rm H}i}({\bm q})
M_{{\rm H}i{\rm O}i}({\bm q})}\right].
\tag{5.39}
\end{align}
Note that the term $M_{\alpha i\alpha i}$ consists of the  $\alpha$th 
individual phonon frequency and the frequency shift due to the coupling 
of particles with  different masses, whereas $M_{\alpha i\beta i}$ plays
a role in the mixing between the individual phonon frequencies.

Next we investigate the phonon frequencies in the two limiting cases:
First we introduce
\begin{align}
v_{\sigma \alpha}^{(n)} & \equiv \frac1{M_\alpha}
\int d^3Rg_{\alpha \alpha}(R)
{V_{\sigma \alpha\alpha} \over 3}\xi^n, 
& v_{\pi \alpha}^{(n)} & \equiv \frac1{M_\alpha}
\int d^3Rg_{\alpha \alpha}(R)
{2V_{\pi \alpha\alpha} \over 3}\xi^n,  \notag \\
{v'}_{\sigma \alpha}^{(n)} & \equiv \frac1{M_\alpha}
\int d^3Rg_{\alpha \beta}(R)
{V_{\sigma \alpha\beta} \over 3}\xi^n, 
& {v'}_{\pi \alpha}^{(n)} & \equiv \frac1{M_\alpha}
\int d^3Rg_{\alpha \beta}(R)
{2V_{\pi \alpha\beta} \over 3}\xi^n,
\tag{5.40} 
\end{align}
$$
c_\alpha \equiv \frac1{M_\alpha} e^2_\alpha
\rho_\alpha\frac1{3\epsilon_0 v_a},
\eqno(5.41)
$$

(i) $qR=\xi \ll 1$

If we define as
\begin{align}
A_{\alpha t} \equiv A_{\alpha x}& =A_{\alpha y}
\equiv {v'}_{\sigma \alpha}^{(0)} +{v'}_{\pi \alpha}^{(0)}
-c_\alpha \equiv A_\alpha -c_\alpha,
& B_{\alpha t}\equiv B_{\alpha x}&=B_{\alpha y}
\equiv \frac{v_{\sigma \alpha}^{(2)}}{10}
+\frac{v_{\pi \alpha}^{(2)}}5,  \notag \\
A_{\alpha l}\equiv A_{\alpha z} 
& \equiv {v'}_{\sigma \alpha}^{(0)} +{v'}_{\pi \alpha}^{(0)}
+2c_\alpha \equiv A_\alpha+2c_\alpha,
& B_{\alpha l} \equiv B_{\alpha z}  &
\equiv \frac{3v_{\sigma \alpha}^{(2)}}{10}
+\frac{v_{\pi \alpha}^{(2)}}{10},  \notag \\
{B'}_{\alpha t}\equiv {B'}_{\alpha x}&={B'}_{\alpha y} 
\equiv \frac{{v'}_{\sigma \alpha}^{(2)}}{10}
+\frac{{v'}_{\pi \alpha}^{(2)}}5,  
& {B'}_{\alpha l} \equiv {B'}_{\alpha z}  
&\equiv \frac{3{v'}_{\sigma \alpha}^{(2)}}{10}
+\frac{{v'}_{\pi \alpha}^{(2)}}{10},  
\tag{5.42}
\end{align}
we obtain
\begin{align}
M_{\alpha x \alpha x} &=M_{\alpha y \alpha y}
  =A_{\alpha t}+B_{\alpha t}= A_\alpha +B_{\alpha t}
-c_\alpha  \notag \\
M_{\alpha z \alpha z} &=A_{\alpha l}+B_{\alpha l}=
A_\alpha+B_{\alpha l}+2c_\alpha, \notag \\
M_{\alpha x \alpha x} &=M_{\alpha y \alpha y}
= \frac{\zeta_\beta}{\zeta_\alpha} \sqrt{\frac{N_\alpha}{N_\beta}}
\{-A_{\alpha t} +{B'}_{\alpha t} \}
=\frac{\zeta_\beta}{\zeta_\alpha} \sqrt{\frac{N_\alpha}{N_\beta}}
\{-A_{\alpha } +{B'}_{\alpha t}+c_\alpha  \}, \notag \\
M_{\alpha z \alpha z} &
= \frac{\zeta_\beta}{\zeta_\alpha} \sqrt{\frac{N_\alpha}{N_\beta}}
\{-A_{\alpha l} +{B'}_{\alpha l} \}
=\frac{\zeta_\beta}{\zeta_\alpha} \sqrt{\frac{N_\alpha}{N_\beta}}
\{-A_{\alpha } +{B'}_{\alpha t}-2c_\alpha  \}.
\tag{5.43}
\end{align}
In the long wavelength limit, we obtain the phonon frequencies:
\begin{align}
q_0^2 &= \omega_{t \pm}^2({\bm q}) \notag \\ 
     &   =\begin{cases} 
             &A_{\rm O}+A_{\rm H}-c_{\rm O}-c_{\rm H} +\frac{2A_{\rm O}-
2c_{\rm O}}{A_{\rm O}+A_{\rm H}-c_{\rm O}-
   c_{\rm H}} (B_{{\rm O}t}-{B'}_{{\rm H}t})
                         +\frac{2A_{\rm H}-2c_{\rm H}}{ A_{\rm O}+
A_{\rm H}-c_{\rm O}-c_{\rm H}} (B_{{\rm H}t}-{B'}_{{\rm O}t}), \\    
             &\frac{2A_{\rm O}-2c_{\rm O}}{ A_{\rm O}+
A_{\rm H}-c_{\rm O}-c_{\rm H}} (B_{{\rm H}t}+{B'}_{{\rm H}t})
                          +\frac{2A_{\rm H}-2c_{\rm H}}{ A_{\rm O}+
A_{\rm H}-c_{\rm O}-c_{\rm H}} (B_{{\rm O}t}+{B'}_{{\rm O}t}).    
            \end{cases}
\tag{5.44}
\end{align}
\begin{align}
q_0^2 &= \omega_{l\pm}({\bm q}) \notag \\
            &=\begin{cases} 
             & A_{\rm O}+A_{\rm H}+2c_{\rm O}+2c_{\rm H} +
\frac{2A_{\rm O}+4c_{\rm O}}{A_{\rm O}+A_{\rm H}+
2c_{\rm O}+2c_{\rm H} } (B_{{\rm O}l}-{B'}_{{\rm H}l})
                 +\frac{2A_{\rm H}+4c_{\rm H}}{ A_{\rm O}+A_{\rm H}+
2c_{\rm O}+2c_{\rm H}} (B_{{\rm H}l}-{B'}_{{\rm O}l}), \\    
             &\frac{2A_{\rm O}+4c_{\rm O}}{ A_{\rm O}+A_{\rm H}+
2c_{\rm O}+2c_{\rm H}} (B_{{\rm H}l}+{B'}_{{\rm H}l})
                 +\frac{2A_{\rm H}+4c_{\rm H}}{ A_{\rm O}+A_{\rm H}+
2c_{\rm O}+2c_{\rm H}} (B_{{\rm O}l}+{B'}_{{\rm O}l}).    
            \end{cases}
\tag{5.45}
\end{align}
$\omega_{\lambda q+}$  are the optical modes and
$\omega_{\lambda q-}$ the acoustic modes.
The velocities of the acoustic phonons are given by 
$$
v_{t}=\sqrt{ \frac{2A_{\rm O}-2c_{\rm O}}{ A_{\rm O}+A_{\rm H}-
c_{\rm O}-c_{\rm H}} (B_{{\rm H}t}+{B'}_{{\rm H}t})
         +\frac{2A_{\rm H}-2c_{\rm H}}
{ A_{\rm O}+A_{\rm H}-c_{\rm O}-
c_{\rm H}} (B_{{\rm O}t}+{B'}_{{\rm O}t}) }\frac1q
\eqno(5,46)
$$
$$
v_{l}=\sqrt{ \frac{2A_{\rm O}+4c_{\rm O}}{ A_{\rm O}+A_{\rm H}+
2c_{\rm O}+2c_{\rm H}} (B_{{\rm H}l}+{B'}_{{\rm H}l})
         +\frac{2A_{\rm H}+4c_{\rm H}}{ A_{\rm O}+A_{\rm H}+
2c_{\rm O}+2c_{\rm H}} (B_{{\rm O}l}+{B'}_{{\rm O}l}) }\frac1q
\eqno(5,47)
$$

(ii) $qR \gg 1$:

In the short wavelength limit, we obtain
\begin{align}
M_{\alpha x \alpha x} &= M_{\alpha y \alpha y}  =
v^{(0)}_{\sigma \alpha}+v^{(0)}_{\pi \alpha}+
{v'}^{(0)}_{\sigma \alpha}+{v'}^{(0)}_{\pi \alpha}  -c_\alpha
\equiv S_{\alpha }-c_\alpha , \notag \\
M_{\alpha z \alpha z} &=
v^{(0)}_{\sigma \alpha}+v^{(0)}_{\pi \alpha}+
{v'}^{(0)}_{\sigma \alpha}+{v'}^{(0)}_{\pi \alpha}+2c_\alpha, 
\equiv S_{\alpha }+2c_\alpha \notag \\
M_{\alpha  x \beta x} &=M_{\alpha y \beta y} =
\frac{\zeta_\beta}{\zeta_\alpha}
\sqrt{\frac{N_\alpha}{N_\beta}}c_\alpha
\notag \\
M_{\alpha z \beta z} &= 
\frac{\zeta_\beta}{\zeta_\alpha}
\sqrt{\frac{N_\alpha}{N_\beta}}(-2c_\alpha),
\tag{5.48}
\end{align}
Then we obtain
$$
q^2_0 =\omega_{t\pm}({\bm q})           
         =\frac12\{S_{\rm O}+S_{\rm H}-c_{\rm O}-c_{\rm H} 
\pm \sqrt{(S_{\rm O}-S_{\rm H}-c_{\rm O}+
c_{\rm H})^2+4c_{\rm O}c_{\rm H}}\},
\eqno(5.49)
$$
$$
q^2_0 =\omega_{l\pm}({\bm q})
            =\frac12\{S_{\rm O}+S_{\rm H}+2c_{\rm O}+2c_{\rm H}
 \pm \sqrt{(S_{\rm O}-S_{\rm H}+2c_{\rm O}-
2c_{\rm H})^2+16c_{\rm O}c_{\rm H}}\},
\eqno(5.50)
$$
These modes correspond to the boson peaks, where 
the density of states has peaks. In nonpolarized liquids, 
because of $c_\alpha =0$ the corresponding logitudinal and tranverse
frequencies and the boson peaks are of the same.

Next we investigate the secular equation (5.38) in the low frequency regime:

(II) $q_0 \tau_{\alpha {\rm M}} \ll 1$

Eq.(5.38) leads to
$$
(q_0+{\rm i}\tau_{\alpha {\rm M}}M_{\alpha i \alpha i})
(q_0+{\rm i}\tau_{\beta {\rm M}}M_{\beta i \beta i})
-{\rm i}\tau_{\alpha {\rm M}}M_{\alpha i \beta i}
 {\rm i}\tau_{\beta {\rm M}} M_{\beta i \alpha i}=0.
\eqno(5.51)
$$
Thus we obtain the dissipative modes due to viscosity
\begin{align}
q_0& =\frac{\rm i}2[\tau_{{\rm OM}}M_{{\rm O}i{\rm O}i}+
\tau_{{\rm HM}}M_{{\rm H}i{\rm H}i}-
\sqrt{(\tau_{{\rm OM}}M_{{\rm O}i{\rm O}i}
-\tau_{{\rm HM}}M_{{\rm H}i{\rm H}i})^2+
4\tau_{{\rm OM}}M_{{\rm O}i{\rm H}i}
\tau_{{\rm HM}}M_{{\rm H}i{\rm O}i} }], \notag \\
&\cong {\rm i}\frac{2\tau_{{\rm OM}}A_{{\rm O}i}}
{ \tau_{{\rm OM}}A_{{\rm O}i}+\tau_{{\rm HM}}A_{{\rm H}i}}
\tau_{{\rm HM}}(B_{{\rm H}i}+{B'}_{{\rm H}i})
         +{\rm i}\frac{2\tau_{{\rm HM}}A_{{\rm H}i}}
{ \tau_{{\rm OM}}A_{{\rm O}i}+\tau_{{\rm HM}}A_{{\rm H}i}}
 \tau_{{\rm OM}}(B_{{\rm O}i}+{B'}_{{\rm O}i}),
\tag{5.52}
\end{align}
where the modes corresponding to the optical modes
disappear because of
the large imaginary value. We should not confuse
the imaginary number ${\rm i}$ and
the suffices $i$; the ${\rm i}$ in front of the terms
means the imaginary number,
but $i$ in the suffices means the component of the coordinates.
Thus viscosities are given by
$$
\eta_t =\sqrt{ \tfrac{2\tau_{{\rm OM}}(A_{\rm O}-c_{\rm O})}
{ \tau_{{\rm OM}}(A_{\rm O}-c_{\rm O})+
\tau_{{\rm HM}}(A_{\rm H}-c_{\rm H})}
\tau_{{\rm HM}}(B_{{\rm H}t}+{B'}_{{\rm H}t})
         +\tfrac{2\tau_{{\rm HM}}(A_{\rm H}-c_{\rm H})}
{ \tau_{{\rm OM}}(A_{\rm O}-c_{\rm O})+
\tau_{{\rm HM}}(A_{\rm H}-c_{\rm H})}
 \tau_{{\rm OM}}(B_{{\rm O}t}+{B'}_{{\rm O}t}) }\frac1q,
\eqno(5.53)
$$
$$
\eta_l =\sqrt{ \tfrac{2\tau_{{\rm OM}}(A_{\rm O}+
2c_{\rm O})}{ \tau_{{\rm OM}}(A_{\rm O}+2c_{\rm O})
+\tau_{{\rm HM}}(A_{\rm H}+2c_{\rm H})}
\tau_{{\rm HM}}(B_{{\rm H}l}+{B'}_{{\rm H}l})
         +\tfrac{2\tau_{{\rm HM}}(A_{\rm H}+
2c_{\rm H})}{ \tau_{{\rm OM}}(A_{\rm O}+2c_{\rm O})+
\tau_{{\rm HM}}(A_{\rm H}+2c_{\rm H})}
 \tau_{{\rm OM}}(B_{{\rm O}l}+{B'}_{{\rm O}l}) }\frac1q,
\eqno(5.54)
$$
where $\eta_l$ and $\eta_t$ correspond to longitudinal and transverse
acoustic phonons, respectively.

\section{The Kauzmann paradox and the VTF law; specific
heat, relaxation times and transport coefficients}
\hspace*{5ex}First, we investigate the entropy due to intra-band density
fluctuations.
Since sound depends on the temperature, we must start with the
interaction Hamiltonian. Here we consider the thermodynamical
function. We introduce the Hamiltonian with
a parameter $\lambda$ \cite{K1,Abr,Fet}:
$$
H(\lambda)=H_0+\lambda H_I.
\eqno(6.1)
$$
The thermodynamical function $\Omega_\lambda$ for the above Hamiltonian
is given by
$$
\Omega_\lambda =-\frac1\beta 
\ln \{{\rm Tr}e^{-\beta(H_0-\mu N+\lambda 
H_I)}\}.
\eqno(6.2)
$$
Here we calculate the thermodynamical function due to intra-band density 
fluctuations denoted by $\Omega_{{\rm intra}\lambda}$.  From Eq.(2.13),
we obtain
\begin{align}
{\partial \Omega_{{\rm intra}\lambda}
\over \partial \lambda}
  &= \frac1\lambda
<\lambda H_\lambda>_\lambda
\cong \frac12\sum_{{\bm q}}
 \frac1{\sqrt{N_\alpha N_\beta}}
V_{\alpha \beta}(q)<\rho^\dagger_{\alpha0{\bm q}}
\rho_{\beta0{\bm q}}>_\lambda, \notag \\
&=-\frac1{2\beta} \sum_{\alpha\beta{\bm q}} 
V_{\alpha \beta}({ q}) 
F^\lambda_{\beta \alpha}(q),
\tag{6.3}
\end{align}
where $<\cdots>_\lambda$ and $F^\lambda(q)$ are calculated by 
$H_I$ replaced by
$\lambda H_I$ in $<\cdots>_c $ and $F(q)$.
Replacing  $V_{\alpha \beta}(q)$ by
$\lambda V_{\alpha \beta}(q)$ into
Eq.(6.3), neglecting the $\lambda$-dependence
of $P_\alpha (q)$ because of
the condition $\beta\hbar |  J_{\mu} | \ll 1$ in Eq.(4.3),
we obtain
$$
 \sum_\gamma\{\delta_{\alpha \gamma}-P_\alpha(q)
\lambda V_{\alpha \gamma}(q)\}
F_{\gamma \beta}^\lambda (q)
=P_\alpha (q)\delta_{\alpha \beta}.
\eqno(6.4)
$$
Eq.(6.4) leads to
$$
\begin{pmatrix}
F^\lambda_{\rm OO} & F^\lambda_{\rm OH} \\
F^\lambda_{\rm HO} & F^\lambda_{\rm HH}
\end{pmatrix}
=\frac1{\det |\quad |}
\begin{pmatrix}
(1-\lambda P_{\rm H} V_{\rm HH})P_{\rm O} & 
\lambda P_{\rm O} V_{\rm OH} P_{\rm H} \\
\lambda P_{\rm H} V_{\rm HO} P_{\rm H}
& (1-\lambda P_{\rm O}V_{\rm OO})P_{\rm H}
\end{pmatrix},
\eqno(6.5)
$$
$$
\det |\quad|=(1-\lambda P_{\rm O} V_{\rm OO})
(1-\lambda P_{\rm H} V_{\rm HH})
-\lambda^2P_{\rm O} V_{\rm OH}P_{\rm H} V_{\rm HO}.
\eqno(6.6)
$$
Thus Eq.(6.3) yields
\begin{align}
{\partial \Omega_{{\rm intra}\lambda} \over \partial \lambda}
&=-\frac1{2\beta}\sum_q 
{(1-\lambda P_{\rm H} V_{\rm HH})P_{\rm O} V_{\rm OO}
+(1-\lambda P_{\rm O} V_{\rm OO})P_{\rm H} V_{\rm HH}
+2\lambda P_{\rm O} P_{\rm H} V_{\rm OH}V_{\rm HO}
\over
(1-\lambda P_{\rm O} V_{\rm OO})
(1-\lambda P_{\rm H} V_{\rm HH})
-\lambda^2P_{\rm O} P_{\rm H} V_{\rm OH}V_{\rm HO} }
\notag \\
&=\frac1{2\beta}\sum_q{\partial \over \partial \lambda}
\ln\{(1-\lambda P_{\rm O} V_{\rm OO})
(1-\lambda P_{\rm H} V_{\rm HH})
-\lambda^2P_{\rm O} P_{\rm H} V_{\rm OH}V_{\rm HO} \}
\tag{6.7}
\end{align}
Integrating Eq.(6.7)
in $\lambda$ and putting $\lambda \rightarrow 1$, we
obtain the thermodynamical function due to the intra-band density 
fluctuations, $\Omega_{\rm intra}$:
$$
\Omega_{\rm intra}=\frac1{2\beta}
\sum_{{\rm i}\nu_n{\bm q}}\ln \{
(1- P_{\rm O} V_{\rm OO})(1- P_{\rm H} V_{\rm HH})
-P_{\rm O} P_{\rm H} V_{\rm OH}V_{\rm HO} \}
+\Omega_{\rm intra 0},
\eqno(6.8)
$$
where the term $\Omega_{\rm intra 0}$ comes from
the initial condition, In order to calculate the entropy
we neglect $1/\tau_0$ terms under the 
condition $\beta\hbar/\tau_0 \ll 1$ and we put
$q_0 \longrightarrow {\rm i}\nu_n$. From Eq.(4.3), 
$P_{\alpha}(q)=\beta (\omega_{ q}^0)^2
/({\rm i}\nu_n^2-(\omega_{\alpha{ q}}^0)^2)$,
we obtain
\begin{align}
\Omega_{\rm intra} &=\frac1{2\beta}\sum_q \ln{
({\rm i}\nu_n^2-\omega^2_{\rm O}({ q}))
({\rm i}\nu_n^2-\omega^2_{\rm H}({ q}))
-\beta^2(\omega^0_{{\rm O}{ q}})^2
(\omega_{{\rm H}{ q}}^0)^2V_{\rm OH}V_{\rm HO}
\over ({\rm i}\nu_n^2-(\omega^0_{{\rm O}{ q}})^2)
({\rm i}\nu_n^2-(\omega^0_{{\rm H}{ q}})^2)}
 + \Omega_{intra0} \notag \\
&=\frac1{2\beta}\sum_q \ln{
({\rm i}\nu_n^2-\omega^2_{\rm s+}({ q}))
({\rm i}\nu_n^2-\omega^2_{\rm s-}({ q}))
\over ({\rm i}\nu_n^2-(\omega^0_{{\rm O}{ q}})^2)
({\rm i}\nu_n^2-(\omega^0_{{\rm H}{ q}})^2)}
 + \Omega_{\rm intra0}
\tag{6.9}
\end{align}

The Contour integration of Eq.(6.9) gives 
$$
\Omega_{\rm intra} =\frac1{\beta}\sum_{\bm q}
[\ln(1-e^{-\beta\hbar\omega_{\rm s+}(q)})
(1-e^{-\beta\hbar\omega_{\rm s-}(q)})
-\ln(1-e^{-\beta\hbar\omega_{{\rm O} q}^0})
(1-e^{-\beta\hbar\omega_{{\rm H} q}^0}) ]
+\Omega_{\rm intra 0}.
\eqno(6.10)
$$
$$
\Omega_{\rm intra0}=
\frac1{\beta}\sum_{\alpha {\bm q}} 
\ln (1-e^{-\beta\hbar\omega_{\alpha{ q}}^0}),
\eqno(6.11)
$$
where we have neglected the term $\frac12\sum_{ q}(\hbar 
\omega_{{\rm s}{ q}}-\hbar \omega_{ q}^0)$.
The first, the second and
the third terms in Eq.(6.10)
correspond to the thermodynamical function
for the sound, the fluctuations due to the bubbles of the intra-band
elementary excitations and the dissipation due to the diffusion,
respectively. The third term compensates the second term and the system
becomes a local equilibrium. From the instability of sound,
Eq.(4.27), the dominant contribution of the temperature dependence
to the thermodynamical function in the first and the second terms
comes from the regions  $q \sim \tilde K$. We separate the region of
wavevector ${\bm q}$ into the number of states
$N_0$ near $\tilde K$ and
the remaining part. Using
 $\beta \hbar \omega_{{\rm s}\pm}(q), \ 
\beta \hbar \omega_{\alpha q}^0 \ll 1$ under 
the condition $\beta\hbar  | J_\mu | \ll 1$,  we obtain
\begin{align}
\Omega_{\rm intra} &\cong \frac{N_0}\beta\ln(
{ \omega_{\rm s+}(\tilde K) \omega_{\rm s-}(\tilde K)
\over  \omega_{{\rm O} q}^0 \omega_{{\rm H} q}^0})
+{\Omega'}_{\rm sound} \notag \\
&=\frac{N_0}\beta\ln\sqrt{
{\omega_{\rm O}(\tilde K)^2
\omega_{\rm H}(\tilde K)^2-\beta^2
(\omega_{{\rm O} q}^0)^2(\omega_{{\rm H} q}^0)^2
V_{\rm OH}(\tilde K)V_{\rm HO}(\tilde K)
\over
(\omega_{{\rm O} q}^0)^2(\omega_{{\rm H} q}^0)^2  }}
 +{\Omega'}_{\rm sound} \notag \\
&=\frac{N_0}{2\beta}\ln\{(1+\beta V_{\rm OO}(\tilde K)
(1+\beta V_{\rm HH}(\tilde K))
-\beta^2V_{\rm OH}(\tilde K)V_{\rm HO}(\tilde K) \}
+{\Omega'}_{\rm sound},
\tag{6.12}
\end{align}
$$
{\Omega'}_{\rm sound}=\frac1\beta{\sum_{\bm q}}'
 [\ln(1-e^{-\beta\hbar \omega_{\rm s+}(q)})
+\ln (1-e^{-\beta\hbar \omega_{\rm s-}(q)}) ],
\eqno(6.13)
$$
where the prime on
$\Omega_{\rm sound}$ and $\Sigma$ means that
the region of the wavevector ${\bm q}$ is limited in the remaining region with
the number of states $N-N_0$ excluded the region with the number of states
$N_0$ near $\tilde K$.
Now we investigate the first term of Eq.(6.12) using Eq.(4.27).
\begin{align}
(1+\beta V_{\rm OO} (\tilde K))&
(1+\beta V_{\rm HH}(\tilde K))
-\beta^2V_{\rm OH}(\tilde K)
V_{\rm HO}(\tilde K)  \notag \\
&=\{1-\frac{T_0}{T}\}\{1+{T_0-(V_{\rm OO}(\tilde K)
+V_{\rm HH}(\tilde K) \over  T} \} \notag \\
&\cong (1-\frac{T_0}{T})(2-{V_{\rm OO}
(\tilde K)+V_{\rm HH}(\tilde K) \over T_0})
\tag{6.14}
\end{align}
Thus we obtain 
$$
\Omega_{\rm intra} \cong 
\frac{N_0}{2\beta}\ln\{1-{T_0 \over T}\} +
{\Omega'}_{\rm sound},
\eqno(6.15)
$$
where we have neglected the constant term, which does not contribute to entropy.
Note that the dissipation compensates the fluctuation entropy in the remaining
region with the number of states $N-N_0$.
The entropy due to the intra-band density fluctuations,
$S_{\rm intra}$
is given by
$$
S_{\rm intra}=S_{\rm K}+S'_{\rm sound},
\eqno(6.16)
$$
$$
S_{\rm K} \cong  - {N_0k_{\rm B} \over 2}\frac{T_0}{T-T_0},
\eqno(6.17)
$$
$$
S'_{\rm sound} \cong -k_{\rm B}{\sum_{\bm q}}'
\{\ln (1-e^{-\beta\hbar\omega_{\rm s+}(q)})
+\ln (1-e^{-\beta\hbar\omega_{\rm s-}(q) }\}+
\frac1T{\sum_{\bm q}}'\{ {\hbar\omega_{\rm s+}( q) \over 
e^{\beta\hbar\omega_{\rm s+}(q)}  -1}
+{\hbar\omega_{\rm s-}( q) \over 
e^{\beta\hbar\omega_{\rm s-}(q)}-1} \}.
\eqno(6.18)
$$
$S_{\rm K}$ manifests the Kauzmann crisis.
We call $S_{\rm K}$ the Kauzmann entropy.

A state of $N$-particles distributed randomly in space corresponds to
a minimum of $N$-particle potential in a configuration space. A hopping
of a particle from a site to a vacancy corresponds to a jump from a
deep valley to another deep valley in the multi-dimentional
configuration space in the energy landscape model (ELM)
as the $\alpha$-relaxation process.
The successive hoppings constitute a
configuration space. A hopping also generates intra-band elementary
fluctuations and the successive hoppings yield the fluctuation
entropy $S_{\rm K}$ due to the intra-band density fluctuations. The hopping
probability is proportional to the configuration number, which is
$e^{S_{\rm K}/k_{\rm B}}$ from the Einstein relation. The hopping probability of a
particle is given uniquely by the hopping amplitude:
$$
J=e^{zS_{\rm K} \over Nk_{\rm B}}=e^{-{E \over T-T_0}},
\eqno(6.19)
$$
where $E=zN_0 T_0/2N$ and $z$ is of the order of the number of the
surrounding H$_2$O molecules.
This equation manifests the Vogel-Tamman-Fulcher (VTF) law. 
The sound $\omega_{\rm s\pm}$  consists of
the individual sound frequencies 
$\omega_{\rm O}(q)$ and $\omega_{\rm H}(q)$.
As seen in the first terms on the right hand side of 
the first line in Eq.(6.12), the mixing between the entropy of sound
$\omega_{\rm s\pm}$ and the fluctuation entropy
due to the intra-band elementary excitations
$\omega^0_{\rm Oq}$ and
$\omega^0_{\rm Hq}$ yields a unique hopping 
amplitude $J$. This fact originates
from the temperature-dependent sound.
Thus  $J$ governs all individual quantities $v_\alpha$ and
$J_{\alpha \mu}$.
Then we can put 
$$
|J_{\alpha \mu}| \propto J,
\eqno(6.20)
$$
$$
\sqrt{U_J^{\alpha \mu\nu}} \propto J.
\eqno(6.21)
$$
Since all $J_{\alpha \mu}$ are governed by a unique $J$,
we can regard $z$ as the order of the number of the surrounding
H$_2$O molecules.
It should be noted that the hopping $J$ generates the intra-band 
elementary excitations, the sound $\omega_{\rm s\pm}(q)$ 
and the Kauzmann entropy $S_{\rm K}$, while
$S_{\rm K}$ determines $J$ in Eq.(6.19).
Thus $\omega_{\rm s \pm}(q)$
in Eq.(4.23) is self-consistently determined.

Next we investigate the entropy due to inter-band density fluctuations.   
The thermodynamical function due to the inter-band density fluctuations
is given in a similar manner to the sound:
\begin{align}
\Omega_{\rm inter} &=\frac1{ 2\beta}\sum_{\lambda {\bm q}}
 \ln{
({\rm i}\nu_n^2-\omega_{\lambda +}^2({\bm q}))
({\rm i}\nu_n^2-\omega_{\lambda -}^2({\bm q})) \over
( {\rm i}\nu_n^2-\omega_{\rm O}^2)
( {\rm i}\nu_n^2-\omega_{\rm H}^2) }
+\Omega_{\rm inter 0},  \notag \\
 & = 
 \frac1{\beta}\sum_{\lambda {\bm q}} 
 [\ln(1-e^{-\beta\hbar\omega_{\lambda +}({\bm q})})
(1-e^{-\beta\hbar\omega_{\lambda -}({\bm q})})
 -\ln(1-e^{-\beta\hbar\omega_{\rm O}})
(1-e^{-\beta\hbar\omega_{\rm H} })]
 +\Omega_{\rm inter 0},
\tag{6.22}
\end{align}
$$
\Omega_{\rm inter 0}=\frac1{\beta}\sum_{\lambda
{\bf q}}\ln(1-e^{-\beta\hbar\omega_{\rm O}})
(1-e^{-\beta\hbar\omega_{\rm H}}) 
\eqno(6.23)
$$
where we have neglected the term 
$\frac12\sum_{\lambda {\bm q}}
(\hbar\omega_{\lambda q}-\hbar\omega)$  																   .
The entropy due to the inter-band density fluctuations is given by
$$
S_{\rm inter}=S_{\rm phonon}+S_\Omega -S_\Omega,
\qquad S_\Omega=S_{\Omega_{\rm O}}+S_{\Omega_{\rm H}},
\eqno(6.24)
$$
$$
S_{\rm phonon}=-k_{\rm B} \sum_{\lambda {\bm q}}
[\ln(1- e^{-\beta\hbar\omega_{\lambda +}({\bm q})})+
\ln(1- e^{-\beta\hbar\omega_{\lambda -}({\bm q})})]
+\frac1{T}\sum_{\lambda {\bm q}}
[{\hbar\omega_{\lambda +}({\bm q}) \over 
e^{\beta \hbar \omega_{\lambda +}({\bm q})}-1}
+{\hbar\omega_{\lambda -}({\bm q}) \over 
e^{\beta \hbar \omega_{\lambda -}({\bm q})}-1}].
\eqno(6.25)
$$
$$
S_{\Omega_\alpha}=k_{\rm B} \sum_{\lambda {\bm q}}
\ln(1- e^{-\beta\hbar\omega_\alpha})
-\frac1{T}\sum_{\lambda {\bm q}}
{\hbar\omega_\alpha \over e^{\beta \hbar \omega_\alpha}-1}.
\eqno(6.26)
$$
In Eq.(6.24), the first, the second and the third terms correspond to
the entropies for the phonons, the fluctuations due to the bubbles of
inter-band elementary fluctuations and
the dissipation due to the viscosity,
respectively. In this case compared with the Kauzmann entropy,
the phonon entropy and the fluctuation entropy do not mix. This fact
originates from the temperature-independent phonons.
The fluctuation entropy lowers
the equilibrium entropy, but the dissipative entropy compensates the
fluctuation entropy and the system becomes to a local equilibrium.

A propagation of an up and down transition
at a site to a surrounding site corresponds to a jump of a shallow valley
to another shallow valley in the multi-dimentional
configuration space in the ELM as the $\beta$-relaxation
process. The successive propagations constitute
another configuration space different from hoppings.
The probability of the magnitude of the randomness of harmonic
frequencies of $\alpha$-particles is proportional to the configuration
number $e^{S_{\Omega_\alpha} /k_{\rm B}}$.
The probability of the magnitude of randomness of
the harmonic frequency of an $\alpha$-particle is proportional to	
$$
\Omega_\alpha=
e^{z_\alpha S_{\Omega_\alpha}/Nk_{\rm B}}=\exp 
\{-{3z_\alpha \beta\hbar \omega_\alpha
\over e^{\beta\hbar\omega_\alpha}-1} \},
\eqno(6.27)
$$
where $z_\alpha$ is  of the order of the
$\alpha$-surrounding particles
of an $\alpha$-particles.
Thus we obtain
$$
\sqrt{U^{\alpha \mu\nu}_{\omega}}\propto \Omega_\alpha.
\eqno(6.28)
$$ 

The Kauzmann entropy $S_{\rm K}$ diverges negatively at $T_0$
so that the system seems to occur the entropy crisis. But the inter-band 
fluctuation entropy $S_{\Omega}$ crosses $S_{\rm K}$  above $T_0$
and prevents the crisis.
The Kauzmann entropy $S_{\rm K}$ dominates above the crossover
temperature, but the fluctuation entropy $S_\Omega$ dominates below the
crossover temperature. But below the crossover temperature, the 
dissipative entropy compensates the $S_{\Omega}$ completely.
We can identify the crossover temperature with
the liquid-glass transition temperature $T_g$:
$$
\left. S_{\rm K}=S_{\Omega} \right|_{T=T_g}.
\eqno(6.29)
$$
Sound is a collision wave essential in a fluid, while phonons are elastic
waves essential in a solid. The glass transion is a sort of dynamical
transition.

Next we investigate the specific heat. First, we calculate the specific
heat due to the entropy of the intra-band density fluctuations, which
consists of the
Kauzmann entropy $S_{\rm K}$ and the entropy due to
the conventional sound, $S'_{\rm sound}$.
The specific heat due to $S_{\rm K}$,
 $C_{\rm K}$ is given by
$$
C_{\rm K}=\left\{ \begin{array}{cr}
{N_0k_{\rm B}T \over 2}
{T_0 \over (T-T_0)^2} & \mbox{ for $T_g<T$} \\
0 & \mbox{for $T<T_g$}
\end{array} \right..
\eqno(6.30)
$$
The specific heat due to the sound, $C'_{sound}$ is given by
$$
C'_{\rm sound}={k_{\rm B} \over 4} {\sum_{\bm q}}'
[ {(\beta\hbar \omega_{\rm s+}(q))^2 \over
\sinh^2 {\beta\hbar\omega_{\rm s+}(q) \over 2}}+
{(\beta\hbar \omega_{\rm s-}(q))^2 \over
\sinh^2 {\beta\hbar\omega_{\rm s-}(q) \over 2}}] 
\cong k_B(N-N_0),
\eqno(6.31)
$$
where we have used the relation
$\beta\hbar\omega_{\rm s \pm}(q) \ll 1$.
$C'_{\rm sound}$ remains $k_{\rm B}(N-N_0)$: constant around
the liquid-glass transition$T_g$. Eqs.(6.30,31)
shows the gap of specific heat at $T_g$.

Next we investigate the specific heat due to the inter-band density
fluctuations. In this case, since the 
inter-band fluctuation entropy cancels with the dissipative entropy,
there remains only the specific heat of phonons. This is because the
structure of the phonons does change little above and below the 
glass transition, while the structure of the sound essentially depends
on the temperature.  We
obtain the conventional specific heat of phonons,
$C_{\rm phonon}$ as
$$
C_{\rm phonon}={k_{\rm B} \over 4}
\sum_{\lambda {\bm q}} 
[{(\beta\hbar \omega_{\lambda +}({\bm q}))^2
\over \sinh^2{\beta\hbar\omega_{\lambda + }
({\bm q}) \over 2}}+
{(\beta\hbar \omega_{\lambda -}({\bm q}))^2
\over \sinh^2{\beta\hbar\omega_{\lambda - }
({\bm q}) \over 2}}].
\eqno(6.32)
$$
At low temperatures, $C_{\rm phonon}\propto T^3$.

Now, we investigate the VTF law on the transport
coefficients, the relaxation times and the velocities of the modes.
  
(i) $T_0<T$: 

The velocities of the particles and the sound only
depend on $J_{\alpha 0}$, so we obtain
$$
v_{\alpha {\rm p}},\ v_{\rm s \pm} \propto J,
\qquad v_{\lambda \pm} \cong {\rm constant}.
\eqno(6.33)
$$
Here it should be noted the phonon velocities
 $v_{\lambda\pm}$ is constant. 

(ii) $T_0<T_g<T$: 

The term $\sqrt{U^{\alpha \mu\nu}}
\propto J_{\alpha \mu}$ dominates; the
$\alpha$-relaxation. We obtain 
$$
\tau_{\alpha0}^{-1}, \ \tau_{\alpha {\rm M}}^{-1} \propto J
\eqno(6.34)
$$
If we consider Eqs.(6.33), (6.34) and diffusivity  in (4.30) and
viscosity $\eta_\lambda$ in Eqs.(5.53) and (5.54),
we obtain
$$
D^\pm, \ \eta_\lambda^{-1} \propto J.
\eqno(6.35)
$$
It should be noted that the sound velocity
$v_{\rm s\pm} \propto J$ in Eq.(6.33)
plays an essential role in the diffusivity $D^\pm$.
Eq.(6.35) satisfies the Stokes
law. 

(iii) $T_0<T<T_g$:

The term $\sqrt{U^{\alpha \mu\nu}_\omega}
\propto \Omega_\alpha$
dominates; the $\beta$-relaxation. Thus we obtain
$$
\tau_{\alpha 0}^{-1},
\  \tau_{\alpha {\rm M}}^{-1} \propto \Omega,
\eqno(6.36)
$$
$$
D^\pm \propto J^2,
\qquad \eta_\lambda \propto \Omega^{-1}.
\eqno(6.37)
$$
In this regime there remains little diffusion.

\section{Concluding Remarks}
\hspace{5ex}
We have calculated the sound and
the diffusion from  intra-band
density fluctuations, and the
phonons and the viscosity from 
inter-band density fluctuations in
water taking into account of
the interaction potentials
$V_{\rm OO}$, $V_{\rm HH}$
and $V_{\rm OH}$, where we have
included the pair distribution
functions $g_{\rm OO}, \  g_{\rm HH},$ and
 $ g_{\rm OH}$.
The terms $V_{\rm OH}$ consists of
$V_{\rm OH}^{\rm hcb}$
and $V_{\rm OH}^{\rm hb}$, where  the pair
distribution function of 
which are given by $g_{\rm OH}^{\rm hcb}$ and
$g_{\rm OH}^{\rm hb}$, respectively.
In water the hydrogen bonding plays an
essential role in constructiong the hydrogen
bonding network. But in the nearest
neighbour approximation the 
$<V_{\rm OH}V_{\rm HO}>_c$
includes the  unphysical terms such as
$V_{\rm OH}^{\rm hcb}
V_{\rm HO}^{\rm hcb}$ and $V_{\rm OH}^{\rm hb}
V_{\rm HO}^{\rm hb}$. Since we concentrate
ourselves to the hydrogen bonding network,
we have taken into account of
only the physical terms,
$V_{\rm OH}^{\rm hcb}V_{\rm HO}^{\rm hb}$ and
$V_{\rm OH}^{\rm hb}V_{\rm HO}^{\rm hcb}$,
 having comfirmed
that when the term $V_{\rm OH}V_{\rm HO}$ appears, 
$V_{\rm OH}V_{\rm HO}$ should be replaced by
$V_{\rm OH}^{\rm hbc}V_{\rm HO}^{\rm hc}+
V_{\rm OH}^{\rm hc}V_{\rm HO}^{\rm hbc}$.

In water the interaction potential $V_{\rm OH}^{\rm hb}$
and the pair distribution function $g_{\rm OH}^{\rm hb}$
are the most important. 
The peak of  $g_{\rm OH}^{\rm hb}$
relates to the hydrogen bonding,
which clusters  water molecules. Thus the sound
instability occurs at a reciprocal particle
distance corresponding
to a hydrogen bond length
$\tilde K \cong \tilde K_{\rm OH}^{\rm hb}$.
If the magnitude of
$V_{\rm OH}^{\rm hb}(a_{\rm OH}^{\rm hb})$ is
large enough compared with the magnitude
$V_{\rm OO}(a_{\rm OO})$ and
$V_{\rm HH}(a_{\rm HH})$, 
the $\omega_{\rm s-}(q)$ and $v_{\rm s-}$
in Eqs.(4.23),  (4.24),  and
the diffusion coefficient $D^-$ in Eq.(4.30) disappear. 

In the term $V_{\rm HH}$ we have taken into account
of the interaction between hydrogens in different water
molecules so that 
the term relates to the hopping of hydrogens. Since we
neglect the interaction between hydrogens in the same
water molecule counting the number of hydrogens, $2N$,
in phonons we obtain the 3-acoustic
modes, and the 3-optical modes which are doubly
degenerate.  Since the number of hydrogens in a
water molecules is 2, if we include the interaction
between hydrogens in the same water molecule,
the degenerate 3-optical phonon modes split into
6-optical ones. 
  
We have elucidated the Kauzmann paradox
on the entropy crisis which originates from the
sound instability  at a reciprocal particle
distance $\tilde K \cong \tilde K_{\rm OH}^{\rm hb}$
corresponding to a hydrogen bond length  and at an
instability temperature $T_0$.
The Kauzmann entropy $S_{\rm K}$ originates
from the mixing between the sound and the 
intra-band fluctuation entropies which determine a unique hopping
amplitude $J$. $J$ governs the velocities $v_{\alpha}$ and the
hopping amplitudes $J_{\mu \alpha}$ for the individual particles
and sound velocity.  But the relative magnitude of the individual particle
velocities are determined by the coupling strength in Eq.(2.8).
On the other hand, since the mixing between  the phonon and
inter-band fluctuation entropies does not occur, the inter-band
fluctuation entropies for the individual particles $S_{\Omega_\alpha}$
are independent. Thus the inter-band fluctuation
entropy $S_\Omega= S_{\Omega_{\rm O}}+S_{\Omega_{\rm H}}$
prevents the Kauzmann entropy
crisis at $T_g$. Below $T_g$ the remaining intra-band fluctuation
entropy free from the mixing and  the inter-band fluctuation
entropy are  compensated by the dissipative entropy due to
diffusion and viscosity, respectively; there remain the sound entropy
only with the remaining number of states $N-N_0$ free from the mixing,
and the complete phonon entropy.

We have elucidated the VTF law on the relaxation times and
the transport coefficients taking into account of the random
scattering processes due to the random hopping amplitudes and
the random harmonic frequencies. The random hopping amplitudes
determine  the VTF law and the $\alpha$-relaxation, while
the random harmonic frequencies determine the $\beta$-relaxation.

We have calculated phonons and viscosity taking into account of
the electric polarization. Since the electric poralization relates to
the p-states of  particles, the electric polarization  connects
the inter-band density fluctuations and does not connects the
intra-band density fluctuations. The effect of the electric
polarization on phonon modes is to separate the
longitudinal and transverse modes constant values
according to the parameters $c_\alpha$ in Eq.(5.41).
We have also elucidated the boson peaks Eqs.(5.49, 50). 

\vspace{0.3cm}
\noindent
{ \Large \bf Acknowledgements}

\vspace{0.3cm}
I would like to thank Professor Michel Peyrard
for stimulating  my interest to the subject
of water and reading the manuscript critically.


\begin{thebibliography}{99}
\bibitem{Careri}
G. Careri, G. Consolini and F. Bruni,
Solid State Ionics {\bf 125} 257-261 (1999)
\bibitem{MP}
M. Peyrard,
Phys. Rev. E {\bf 64} 011109-1-5 (2001)
\bibitem{Halle}
B. Halle, 
in {\em Hydration processes in biology}, M.-C. Bellissent-Funel (Ed.),
IOS Press 1999, page 233
\bibitem{Russo}
D. Russo, G. Hura and T. Head-Gordon,
Biophysical Journal {\bf 86}, 1852-1862 (2004)
\bibitem{K1}T. Kitamura, Phys. Rep. 383 (2003)1.                       
\bibitem{K2}T. Kitamura, AIP Conference Proceedings {\bf 708} 631-634 (2004).
\bibitem{K3}T.~Kitamura, Phys. Lett. A 147 (1990) 511.	
\bibitem{K4}T.~Kitamura, M.~Silbert, Phys. Lett. A 215 (1996) 69.   
\bibitem{K5}T.~Kitamura, Il Nuovo Cimento D 11 (1989) 1441.
\bibitem{Mar}A.A.~Maradudin, E.W.~Montroll, G.H.~Weiss, I.P.~Ipatova, 
            Solid State	Phys. Supple.3 (2st ed. 1971).
\bibitem{Ewa}P.~P.~Ewald, Ann. Phys. 54 (1917) 519,557; 64 (1921) 253.
\bibitem{K6}T.~Kitamura, Phys. Lett. A {\bf 282}, 59 (2001).
\bibitem{Eis}D.~Eisenberg, W.~Kauzmann, The structure and Properties of Water
            (Oxford, London, 1969).
\bibitem{Han}J.~P.~Hansen and I.~R.~MacDonald, Theory of Simple Liquids
            2nd ed. (Academic Press, London, 1990).		
\bibitem{Hey}D.~M.~Heyes, The Liquid State (Wiley, Chichester, 1997).
\bibitem{Sta}H.~E.~Stanley, in Hydration Processes in Biology, NASO ASI series
                   Vol. 305, M.~C.~Bellissent-Funel ed. (IOS Press, Amsterdam, 1999). p.1.
\bibitem{Abr}A.A.~Abrikosov, L.P.~Gorkov, I.E.~Dzyloshinski, Methods of
            Quantum Field Theory in Statistical Physics ( Prentice-Hall,
            NJ, 1963).		
\bibitem{Fet}A.L.~Fetter, J.D.~Walecka, Quantum Theory of Many-Particle
            Systems (MacGrow-Hill, New York, 1971).		

\end{thebibliography}
\end{document}